\documentclass[sn-mathphys,iicol]{sn-jnl}

\usepackage{graphicx}%
\usepackage{amsmath,amssymb,amsfonts}%
\usepackage{amsthm, mathtools}%
\usepackage{mathrsfs}%
\usepackage[title]{appendix}%
\usepackage{xcolor}%
\usepackage{textcomp}%
\usepackage{manyfoot}%
\usepackage{booktabs}\let\cline\cmidrule%
\usepackage{algorithm}%
\usepackage{algorithmicx}%
\usepackage{algpseudocode}%
\usepackage{listings}%
\usepackage{multicol}
\usepackage{multirow}%
\usepackage{bm}
\usepackage{caption}
\usepackage{subcaption}
\usepackage{enumitem}
\usepackage{stfloats}

\newcommand{\mt}[1]{\mathtt{#1}}
\newcommand{\ttt}[1]{\texttt{#1}}
\newcommand{\mb}[1]{\mathbf{#1}}

\usepackage{siunitx}

\newcommand{\PKnew}[1]{\textcolor{black}{#1}}

\newcommand{\Tcolor}[1]{def}
\definecolor{LightCyan}{rgb}{0.88,1,1}
\definecolor{mygreen}{RGB}{28,172,0} 
\definecolor{mylilas}{RGB}{170,55,241}
\usepackage{tikz,pgfplots}

\hypersetup{pdftex=true, colorlinks=true, breaklinks=true, linkcolor=black, menucolor=black, citecolor=black, urlcolor=black}

\theoremstyle{thmstyleone}%
\theoremstyle{thmstyletwo}%

\theoremstyle{thmstylethree}%

\begin{document}

\title[Article Title]{\texttt{PyTOaCNN}: Topology optimization using an adaptive convolutional neural network  in Python}

\author{\fnm{Khaish} \sur{Singh Chadha}}
\author*{\fnm{Prabhat} \sur{Kumar}}\email{pkumar@mae.iith.ac.in}

\affil{\orgdiv{Department of Mechanical and Aerospace Engineering}, \orgname{Indian Institute of Technology Hyderabad}, \orgaddress{\state{Telangana} \postcode{502285},  \country{India}}}


\abstract{This paper introduces an adaptive convolutional neural network (CNN) architecture capable of automating various topology optimization (TO) problems with diverse underlying physics. The proposed architecture has an encoder-decoder-type structure with dense layers added at the bottleneck region to capture complex geometrical features. The network is trained using datasets obtained by the problem-specific open-source TO codes. Tensorflow and Keras are the main  libraries employed to develop and to train the model. Effectiveness and robustness of the proposed adaptive CNN model are demonstrated through its performance in compliance minimization problems involving constant and design-dependent loads and in addressing bulk modulus optimization. Once trained, the model takes user's input of the volume fraction as an image and instantly generates an output image of optimized design. The proposed CNN produces high-quality results resembling those obtained via open-source TO codes with negligible performance and volume fraction errors. The paper includes complete associated Python code (Appendix~\ref{Sec:PyTOaCNN}) for the proposed CNN architecture and explains each part of the code to facilitate reproducibility and ease of learning.}


\maketitle

\section{Introduction}
Topology optimization (TO), a computational technique, finds the optimum material distribution within a given design domain while extremizing an objective with predetermined constraints and boundary conditions~\citep{sigmund2013topology}. It provides innovative and resource-efficient optimized designs for problems with single- and/or multi-physics. These days, demands for TO methods have increased considerably in academia and industry due to recent developments in additive manufacturing that can print complex geometry obtained from the TO approaches~\citep{bayat2023holistic}. A typical TO  involves four stages: (i) discretizing the design domain using finite elements (FEs). Each element is allotted a design variable $\rho \in [0,\,1]$. $\rho =1$ and $\rho=0$ indicate solid phase and void phase of the element, respectively  (ii) conducting FE analysis (ses) for relevant physical aspects, (iii) determining the objective function, constraints, and their corresponding sensitivities, and (iv) updating the design variable through the optimization procedure. TO can become notably computationally demanding when dealing with large-scale problems~\citep{choi2016comparison} and can become intricate and complex with multi-physics/design-dependent loads~\citep{bayat2023holistic,kumar2020topology,kumar2022topology,siqueira2024topology}.
\begin{figure}
	\centering
	\includegraphics[scale=1]{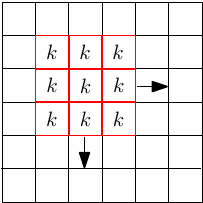}
	\caption{Kernal movement}
	\label{fig:kernel_movement}
\end{figure}

To complement the traditional TO techniques and confront the computational intensity inherent to them, the integration of deep learning into optimization tasks has emerged as a promising avenue~\citep{sosnovik2019neural,banga20183d,sasaki2019topology,harish2020topology,chandrasekhar2021tounn,regenwetter2022deep}. Previous attempts in using deep learning methods to
automate TO problems have already shown potential~\citep{lee2020cnn,chandrasekhar2021tounn,regenwetter2022deep,ramu2022survey}. Convolutional Neural Networks (CNNs) have demonstrated proficiency in extracting valuable features and discerning intricate patterns and relationships within image data~\citep{lee2020cnn,banga20183d,lee2020cnn,ramu2022survey}. TO methods generate visual representations of optimized designs in the form of images; hence, using CNNs to automate TO problems becomes a natural choice.

\cite{sosnovik2019neural} provided a neural network (NN) architecture to improve the optimum material layout for 2D optimization problems. They trained the model using 10,000  training data samples. \cite{banga20183d} proposed an NN architecture for 3D problems, which takes the intermediate solutions of an optimization problem as input to provide the final optimized designs. The method used 6000 training data samples in the study. \cite{oh2019deep} integrated TO and generative adversarial network in their method. \cite{yu2019deep} proposed an approach wherein a CNN architecture is used to obtain low-resolution solutions, and then a conditional generative adversarial network is used to obtain the high-resolution optimized solutions. \cite{zhang2020deep} provided a deep-leaning model capable of automating and producing optimized designs for boundary conditions that it was not trained on. However, initial nodal displacement and strain information of the boundary and loading conditions for which the optimized design will be generated are required. \cite{lee2020cnn} proposed a recognition method based on the CNN model for TO that eliminates the FE analysis step and accelerates the TO process. The model proposed in~\cite{xiang2022accelerated} focused on accelerating the optimization process. \cite{xue2021efficient} provided a deep learning model for achieving high-resolution results. \cite{seo2023topology} proposed a machine-learning-based surrogated model to predict the optimum material layout of the problems using the encoder-decoder network, Unet, and Unet++. A reader can refer to \cite{ramu2022survey,regenwetter2022deep} for a detailed overview of the NN-based TO methods. 
 
CNN architectures developed in the past were tailored for specific optimization tasks and lacked the generalization ability to adapt to new optimization problems for different applications. In addition, they also required a considerable amount of data. Herein, we propose a simple and efficient CNN architecture to automate diverse TO problems requiring minimal sample data, inspired by the encoder-decoder model per~\cite{harish2020topology}. The proposed model is fluidic; it contains the adaptive layer that has a variable number of neurons. This ensures that the user has a certain amount of control over the network output. In addition, the paper constituents with essential concepts necessary for understanding the model's functionality,  provides the complete Python code in Appendix~\ref{Sec:PyTOaCNN}, and furnishes explanations of each part of the code in detail. Codes using machine/deep learning techniques for TO are necessary in academia and the professional landscape. These codes can provide practical gateways and platforms, serve as education tools, and offer hands-on experience for students, researchers, and newcomers to use, learn, develop, and extend machine/deep learning with TO. The code is envisioned to provide an introductory example that facilitates a potential avenue for understanding deep learning techniques with TO. In addition, we outline how one can use the proposed architecture to automate their respective TO problems effectively.

In summary, this paper offers the following new contributions:

\begin{itemize}
	\item Provides an adaptive convolution neural network for topology optimization, which can cater to multidisciplinary design optimization problems having different applications.
	 \item The proposed model is capable of producing high-quality optimized results with a small set of training data. 
     \item The adaptive layer of the model grants users control over the quality of outputs generated by choosing a suitable number of neurons $n$
	 \item The efficacy and success of the proposed model are demonstrated on compliance minimization problems with constant and design-dependent loads, and on maximizing the material bulk modulus.
	 \item The pertinent Python code is provided in Appendix~\ref{Sec:PyTOaCNN} and is explained in detail. We believe the code will assist researchers, newcomers, and students besides reproducing the results.
\end{itemize}

The remainder of the paper is structured as follows: Sec.~\ref{Sec:FundCNN} provides the fundamental of CNNs in brief.   Sec.~\ref{Sec:ProCNN} introduces and explains the proposed CNN architecture. Section \ref{Sec:probdesc} describes different TO problems that are automated for testing accuracy of the proposed model. Sec.~\ref{Sec:GentrainData} explains the methodology for generate training data. Sec.~\ref{Sec:Pyimpl} provides Python code implementation with a detailed explanation. Sec.~\ref{Sec:resndis} presents the optimized results generated by the proposed CNN and compares these results with their counterparts generated via open-source MATLAB codes. Finally, capability of the proposed machine learning model, its limitations, and future scope are summarized in Sec.~\ref{Sec:conc}
\begin{figure*}
	\centering
	\includegraphics[scale=1.5]{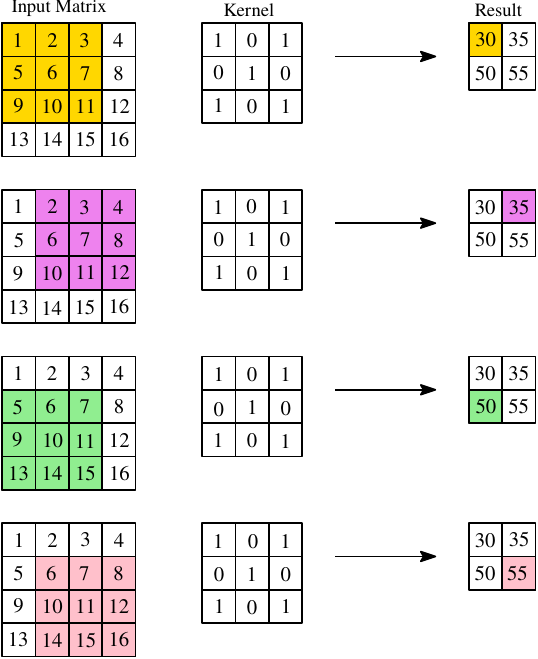}
	\caption{Kernal operation}
	\label{fig:kernel_operation_new}
\end{figure*}
\section{Fundamentals of CNNs}\label{Sec:FundCNN}
This section provides the foundational concepts of CNNs in brief for the sake of completeness. Such a network employs convolution operation, max pooling operation, and transpose convolution operations, which are discussed below. 

\subsection{Convolution operation}\label{Sec:ConOp}

In a CNN, the convolution operation is a detective searching for patterns in an image. Consider an image as a grid of pixels, and the convolutional layer uses small filters (kernels) to slide over this grid (Fig.~\ref{fig:kernel_movement}). At each step, the filter looks at a small portion of the image and checks for specific features, like edges or textures. The results are combined to create a new ``feature map'' that highlights important aspects of the image. This helps the network understand the visual elements in the input data.

A kernel is a small 2D matrix that maps onto the input image by simple matrix multiplication and addition. The output of the convolution operation has lower dimensions and is, therefore, easier to work with. The shape of a kernel is dependent mainly on the input shape of the image and architecture of the entire network primarily; the size of kernels is a square matrix~(Fig.~\ref{fig:kernel_movement}). The movement of a kernel is always from left to right and top to bottom, as displayed in Fig.~\ref{fig:kernel_movement}, which is determined by stride. 

Stride is defined as the step by which the kernel moves. For example, a stride of 1 makes the kernel slide by one row/column at a time. Likewise, a stride of 2 moves the kernel by two rows/columns at a time. In summary, the stride of $n$ makes the kernel slide by $n$ rows/columns at a time.

\begin{figure}
	\centering
	\includegraphics[scale=1]{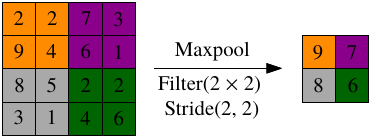}
	\caption{Maxpooling operation}
	\label{fig:maxpooling_operation}
\end{figure}

\begin{figure*}
	\centering
	\includegraphics[scale=1.15]{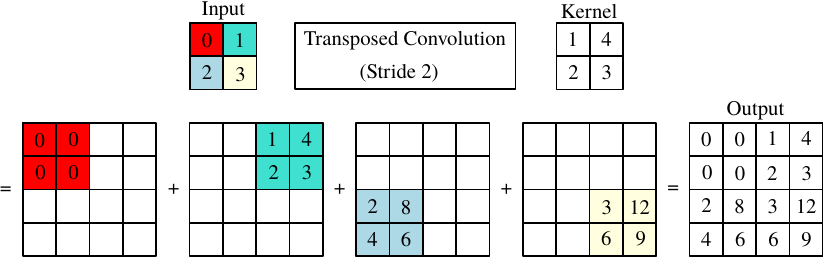}
	\caption{Transposed Convolution Operation}
	\label{fig:transconv_operation}
\end{figure*}
\begin{figure*}
	\centering
	\includegraphics[scale=1]{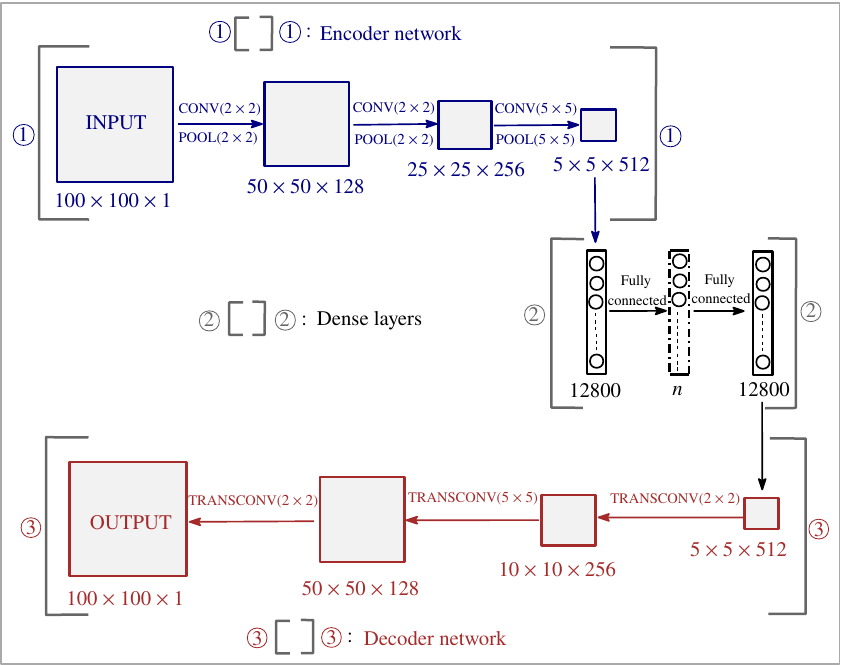}
	\caption{General Architecture of the Adaptive Convolutional Neural Network (CNN)}
	\label{fig:CNNarchitech}
\end{figure*}
In convolution operations, typically, padding is performed to the input image, which involves adding extra pixels (zeros) to avoid information loss at the edges. Depending upon the type of padding used, one can classify convolution operations into two types:
\begin{itemize}
	\item \textit{Same convolution}: It involves padding the image so that the output size is the same as the input size.
	\item \textit{Valid convolution}: It involves no padding on the input data, resulting in a smaller output size
\end{itemize}
We have used the `same' padding in the proposed CNN model (Sec.~\ref{Sec:ProCNN}).
One can calculate the output size $(O)$ of a convolution operation given input size $(I)$, filter size $(K)$, stride $(S)$, and padding $(P)$ as~\citep{Goodfellow-et-al-2016}
\begin{equation}
	O = \frac{I - K + 2P}{S} + 1
\end{equation}
Next, we describe the working of a kernel operation on a sample matrix. We take input matrix of $4\times 4$ and kernel with size $3\times 3$, see Fig.~\ref{fig:kernel_operation_new}. As the shape of the input matrix is larger than that of the kernel, we can implement a sliding window protocol and apply the kernel over the entire input. For example, the first entry in the convoluted result is evaluated as $1\times1 + 2\times0 + 3\times1 + 5\times0 + 6\times1 + 7\times0 + 9\times1 + 10\times0 + 11\times1= 30$ (Fig.~\ref{fig:kernel_operation_new}). One performs the following steps for the sliding  window protocol:
\begin{itemize}
	\item First, the kernel gets into position at the top-left corner of the input matrix (Fig.~\ref{fig:kernel_operation_new}, 1st row, 1st column, 1st image).
	\item Then it starts moving left to right, calculating the dot product and saving it to a new matrix until it has reached the last column (Fig.~\ref{fig:kernel_operation_new}, 2nd row, 1st column, 1st image).
	\item Next, the kernel resets its position at the first column, but now it slides one row to the bottom. Thus following the fashion left-right and top-bottom.
	\item Steps 2 and 3 are repeated till the entire input is processed.
\end{itemize}
A sample for kernel operations performed is shown in Fig.~\ref{fig:kernel_operation_new}. 
\subsection{Maxpooling operation}\label{Sec:MaxPooling}
MaxPooling is a down-sampling operation often used in CNNs to reduce the spatial dimensions of the input volume (Fig.~\ref{fig:maxpooling_operation}). It is a pooling layer that helps retain the most essential information while discarding less critical ones. MaxPooling is typically applied after convolutional layers in a CNN.
The basic idea behind MaxPooling is to divide the input image into non-overlapping rectangular regions and, for each region, output the maximum value. This operation is performed independently for each channel in the input. Typical choices for the size of the pooling window are $2\times2$ or $3\times3$, and the stride (the step size when moving the pooling window) is often set to be equal to the size of the window for non-overlapping pooling.

\subsection{Transpose convolution operation}\label{Sec:TConOp}
Transpose convolution is like an artist enlarging a small painting on a grid. Imagine that each square in the grid represents a pixel in a low-resolution image. Transpose convolution helps recreate a larger version of the painting by placing new pixels strategically (Fig.~\ref{fig:transconv_operation}). It is as if one is zooming into the original artwork, adding details, and making it bigger. This proves highly valuable for reconstructing an image following a sequence of convolution operations that have decreased its dimensions.

The transposed convolutional layer is similar to a standard convolutional layer, except it performs the convolution operation in the opposite direction. Instead of sliding the kernel over the input and performing element-wise multiplication and summation, a transposed convolutional layer slides the input over the kernel. It performs element-wise multiplication and summation, as shown in the example in Fig.~\ref{fig:transconv_operation}. This results in an output larger than the input and the output size can be controlled by the stride and padding parameters of the layer. The output size of the transpose convolution generation $O$ is calculated as~\citep{Goodfellow-et-al-2016}
	\begin{equation}
		O = (I - 1) S + K - 2P
	\end{equation}

\subsection{Convolutional Neural Network}\label{Sec:CNN}
A CNN is formed by stacking multiple convolutional layers for hierarchical feature extraction from input data. The introduced convolutional layers use learnable filters to capture local patterns in the input. CNN architecture learns by obtaining weights and biases that minimize training data errors, wherein weights are the specific elements within kernels, highlighting the concept of learnable filters employed by convolutional layers. In each convolutional layer of a  CNN, after the convolution operation is performed, a bias is added to each output element to introduce an offset, allowing the model to account for any inherent biases in the data. The output is passed through an activation function, which introduces nonlinearity to the model. This nonlinearity is essential for capturing complex structural features and relationships in the data. This empowers CNN to model a broader range of functions. Next, we describe the proposed CNN architecture.
\section{Proposed CNN}\label{Sec:ProCNN}
The proposed architecture has an encoder-decoder-type structure with dense layers added at the bottleneck region. The encoder and decoder parts of the network are based on convolutional. The architecture combines the strengths of convolutional layers for feature extraction from images and fully connected layers (dense layers) for relatively more abstract, high-level processing. The three main parts of the proposed architecture are discussed below.
\begin{description}
	\item \textbf{Encoder network:} 
	It plays a crucial role in extracting meaningful information from the training data while reducing the dimensionality of the input data. In this context, the input image, with dimensions $(100\times100\times1)$, undergoes down-sampling through the encoder network to reach a size of $(5\times5\times512)$. This downsizing process is accomplished through three consecutive convolutional and max-pooling operations (refer to Fig.~\ref{fig:CNNarchitech}). All convolution operations are conducted as "same" convolutions, ensuring that information at the edges of the input image is thoroughly considered in the resulting feature map. The responsibility of image down-sampling lies with the max-pooling operations.
	
	\item \textbf{Dense layers:} The encoder network's output, sized $(5\times5\times512)$, is flattened to form the initial dense layer comprising 12800 neurons (Fig.~\ref{fig:CNNarchitech}). An adaptive (optional) layer (depicted by dotted boundaries) having a variable number of neurons $n$) is placed. Another dense layer with 12800 neurons is followed after that. The inclusion of the adaptive layer endows the network with the ability to automate a diverse array of optimization tasks. Including the adaptive layer, the required number of neurons $n$ is based on the complexity of the optimized design for the particular TO problems.
	
	\item \textbf{Decoder network:} This network uses successive transpose convulution operations (Fig.~\ref{fig:CNNarchitech}) to up-scale the image size to $(100\times100\times1)$. Zero padding for input and output for all the transpose convolution operations is used. One can note the employed filter size and number of filters for each transpose convolution from Fig.~\ref{fig:CNNarchitech}. 
\end{description}
The proposed CNN architecture is termed the \textit{Base} architecture when devoid of the adaptive layer. We use the "ReLU" (Rectified Linear Unit, cf.~\cite{ronneberger2015u}) activation function in all the layers. The Mean Squared Error(MSE) cost function is employed. For efficient training purposes, the "Adam" (Adaptive Moment Estimation, cf.~\cite{kingma2014adam}) optimizer is utilized.

\begin{figure*}[h]
	\begin{subfigure}{0.30\textwidth}
		\centering
		\includegraphics[scale=0.6]{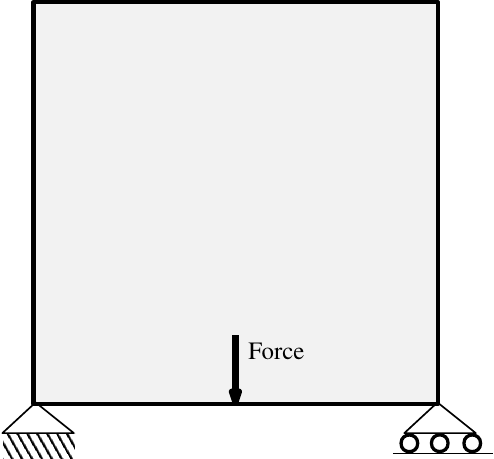}
		\caption{}
		\label{fig:mid_load}
	\end{subfigure}
	\quad
	\begin{subfigure}{0.30\textwidth}
		\centering
		\includegraphics[scale=0.7]{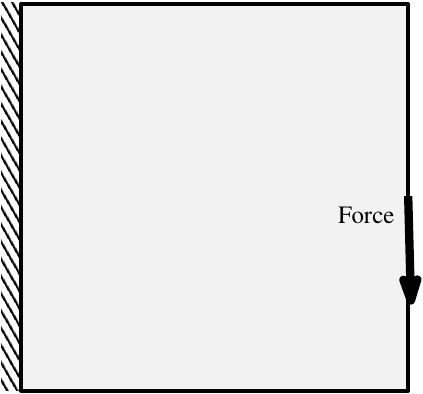}
		\caption{}
		\label{fig:can_1cent}
	\end{subfigure}
\quad
\begin{subfigure}{0.30\textwidth}
	\centering
	\includegraphics[scale=0.7]{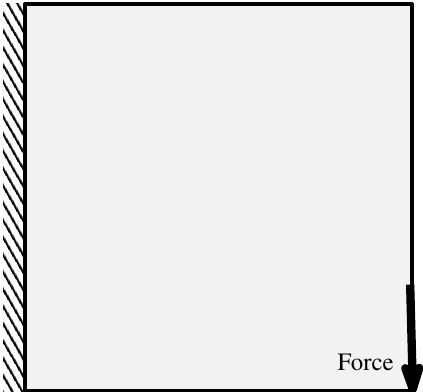}
	\caption{}
	\label{fig:can_1end}
\end{subfigure}
\quad
	\begin{subfigure}{0.45\textwidth}
	\centering
	\includegraphics[scale=0.65]{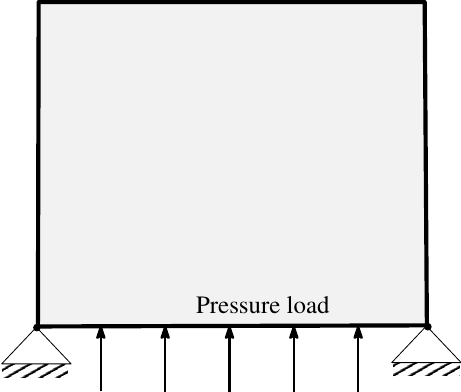}
	\caption{}
	\label{fig:archDD}
\end{subfigure}
\quad
\begin{subfigure}{0.45\textwidth}
	\centering
	\includegraphics[scale=0.65]{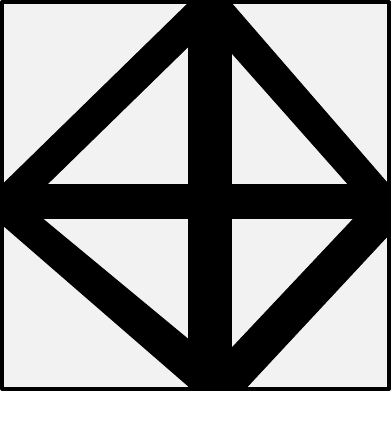}
	\caption{}
	\label{fig:toxDD}
\end{subfigure}
	\caption{Problem description (a) Mid-load (b) Cantilever beam with center load (c) Cantilever beam with end load (d) Pressure loadbearing arch and (e) Bulk modulus maximization} \label{fig:Deasign domains}
\end{figure*}

\section{Problem description and optimization formulation}\label{Sec:probdesc}
To showcase the network's and presented Python code's capability in automating various topology optimization challenges, we evaluate its performance on the compliance minimization problem with constant and design-dependent loads. 

The design domains are parameterized using bi-linear quadrilateral finite elements. Each element is assigned a design variable $\rho$. The modified solid isotropic material with penalization (SIMP) approach~\citep{sigmund2013topology} is implemented to interpolate the Young's modulus of element $i$ as:
\begin{equation}
	E_i = E_0 + \rho_i^p (E_1- E_0)
\end{equation}
where $E_1$ and $E_0$ are the Young's moduli of the solid and void phases of element~$i$, respectively. $p$ is the SIMP parameter. We use $p =3$ for all the examples presented. Next, we describe the optimization formulation for the numerical examples solved in this paper.
\subsection{With constant load}
A constant load does not change its direction and location with the TO progress. Such loads have many applications~\citep{sigmund2013topology}. To determine the optimized designs with constant loads, typically, the following standard compliance minimization problem is solved with a given resource constraint:
\begin{equation} \label{EQ:OPTI} 
	\begin{rcases}
		\begin{split}
			&{\min_{\bm{\rho}}} \quad C({\bm{\rho}}) = \mb{u}^\top \mb{K}(\bm{\rho})\mb{u} \\
			&\text{subjected to:}\\
			&\bm{\lambda}:\,\,\mb{K} \mb{u} - \mb{F} = \mb{0}\\
			&\Lambda:\,\, V(\bm{\rho})-V^* \le 0\\
			&\quad\,\,\,\, \bm{0} \leq \bm{\rho} \leq \bm{1} 
		\end{split}
	\end{rcases},
\end{equation} 
where $C$ denotes the structure's compliance and $nel$ is the total number of FEs used to describe the design domain. $\mb{K}$ and $\mb{u}$ are the global stiffness matrix and global displacement vector, respectively. $\mb{F}$ is the global force vector. $V^*$ and $V$ denote the permitted and current volume of the design domain. $\bm{\lambda}$  and $\Lambda$ are the Lagrange multipliers. $\bm{\rho}$ is the design vector. \ttt{top88} MATLAB code~\citep{andreassen2011efficient} is used to obtain the output data pertaining to this set of problems.

\subsection{With design-dependent load}
A design-dependent load changes its direction and location with the TO progress. Though such loads are prevalent in various applications, they pose several challenges~\citep{kumar2020topology}. Typically, to obtain the pressure load-bearing structures, structure's compliance is minimized with the given volume constraint. We solve the optimization problem described in~\citep{kumar2020topology} for designing the load-bearing structures.  
\begin{equation} \label{Eq:OPTIDD} 
	\begin{rcases}
		\begin{split}
			&{\min_{\bm{\rho}}} \quad C({\bm{\rho}}) = \mb{u}^\top \mb{K}(\bm{\rho})\mb{u} \\
			&\text{subjected to:}\\
			&\qquad\qquad \qquad\bm{\lambda}_1:\,\,\mb{A} \mb{p} = \mb{0}\\
			&\qquad \qquad\qquad\bm{\lambda}_2:\,\,\mb{K} \mb{u} = \mb{F} = -\mb{T}\mb{p}\\
			&\qquad \qquad\qquad\Lambda:V(\bm{\rho})-V^* \le 0\\
			&\qquad \qquad\qquad\quad\,\,\,\, \bm{0} \leq \bm{\rho} \leq \bm{1} 
		\end{split}
	\end{rcases},
\end{equation} 
where $\mb{A}$, $\mb{p}$, and $\mb{T}$ are the global flow matrix, pressure load, and transformation matrix. Readers may refer to \cite{kumar2020topology} for a detailed overview of the problem setting. The method introduced in~\cite{kumar2020topology} has been extended to solve a set of different problems, e.g., 3D structures and compliant mechanisms~\citep{kumar2021topology}, for the prescribed length-scale compliant mechanisms~\citep{kumar2022topological}, soft grippers~\citep{pinskier2024diversity}, a PneuNet of a soft robot~\citep{kumar2022towards}, multi-material frequency contained TO~\citep{banh2024frequency}, to name a few. Using the method reported in~\cite{kumar2020topology}, \ttt{TOPress} MATLAB code is presented in~\cite{kumar2023TOPress}, which is used herein to generate the output data for the training of the proposed CNN model.
\subsection{Material bulk modulus optimization}
We solve the following optimization problem for obtaining the microstrucutre~\citep{xia2015design}

\begin{equation} \label{EQ:OPTIM} 
	\begin{rcases}
		\begin{split}
			&{\min_{\bm{\rho}}} \quad C({E_{ijkl}^H(\bm{\rho})}) \\
			&\text{subjected to:}\\
			&\mb{K} \mb{u}^{A(kl)} - \mb{F}^{kl} = \mb{0}, k,\,l = 1,\cdots,d\\
			& V(\bm{\rho})-V^* \le 0\\
			&\quad\,\,\,\, \bm{0} \leq \bm{\rho} \leq \bm{1} 
		\end{split}
	\end{rcases},
\end{equation} 
where $u^{A(kl)}$ and $\mb{F}^{kl}$ are the global displacement vector and force vector for the test case ($kl$), respectively~\citep{xia2015design}.  $C({E_{ijkl}^H(\bm{\rho})})$ is a function of the homogenized stiffness tensor~\citep{xia2015design}. Herein, we have considered maximization of the material bulk modulus, i.e., $c = (E_{1111} + E_{1122} + E_{2211} + E_{2222})$. \ttt{topX} code provided in~\cite{xia2015design} is employed to obtain the required output data for this case. Readers may refer to~\cite{xia2015design} for more details about the problem framework. 
\begin{figure*}
	\begin{subfigure}{0.30\textwidth}
		\centering
		\includegraphics[scale=0.20]{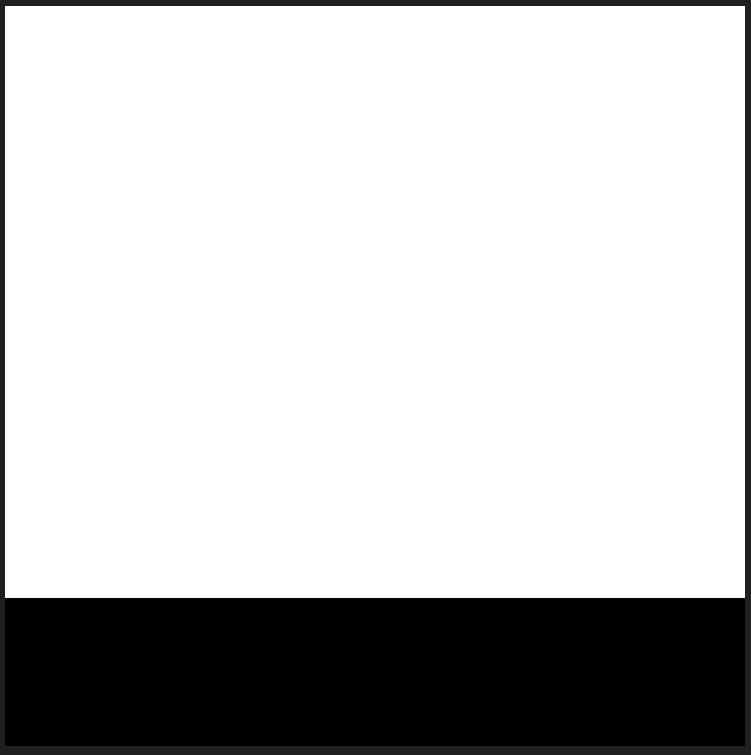}
		\caption{}
		\label{fig:inputimage}
	\end{subfigure}
\quad
\begin{subfigure}{0.30\textwidth}
	\centering
	\includegraphics[scale=0.20]{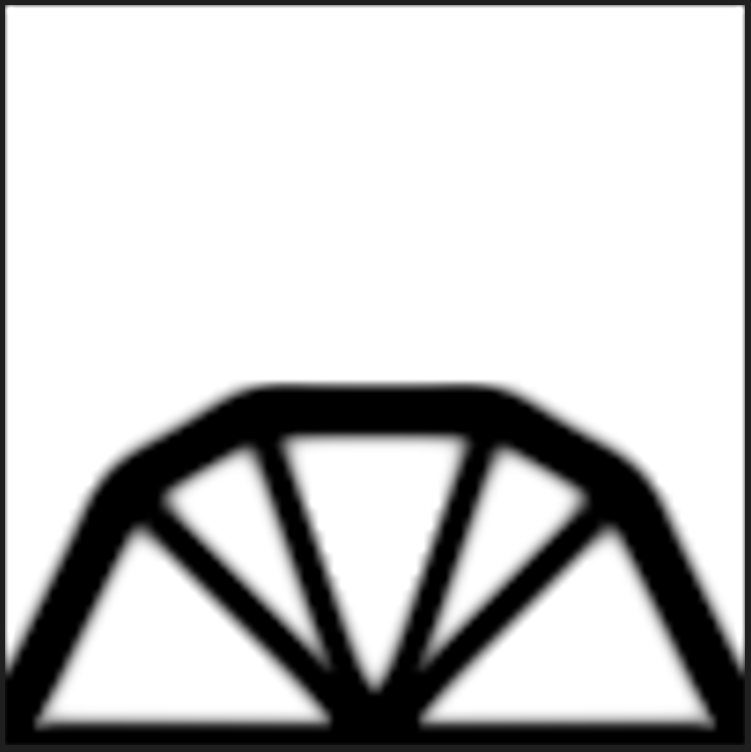}
	\caption{}
	\label{fig:so_ss_full}
\end{subfigure}
\quad
\begin{subfigure}{0.30\textwidth}
	\centering
	\includegraphics[scale=0.20]{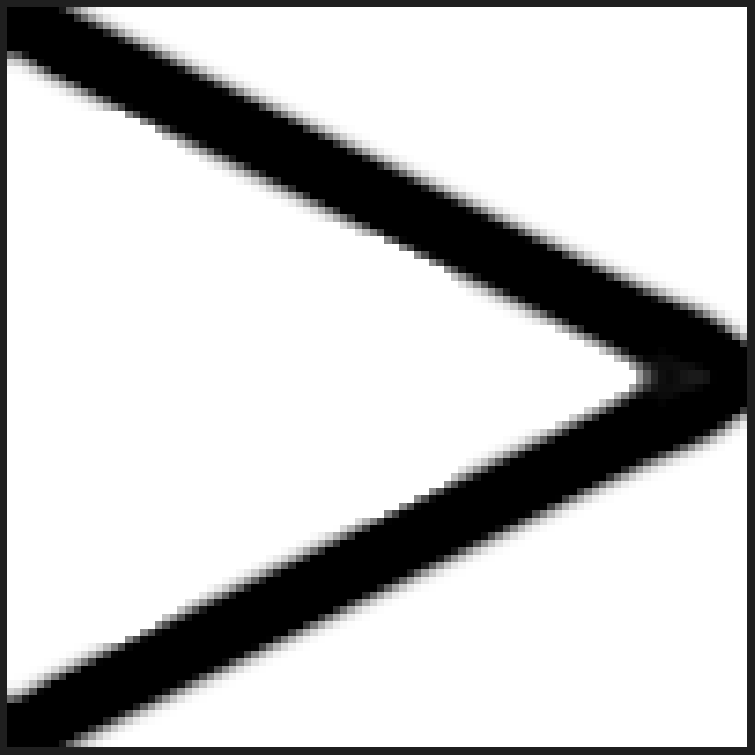}
	\caption{}
	\label{fig:so_x}
\end{subfigure}
	\quad
	\begin{subfigure}{0.30\textwidth}
		\centering
		\includegraphics[scale=0.20]{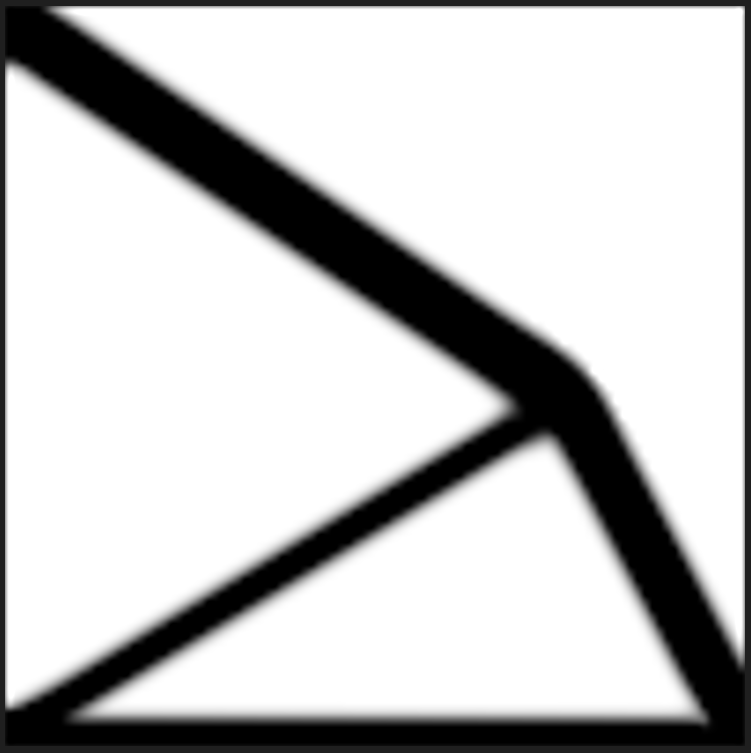}
		\caption{}
		\label{fig:so_bot}
	\end{subfigure}
	\quad
	\begin{subfigure}{0.30\textwidth}
		\centering
		\includegraphics[scale=0.20]{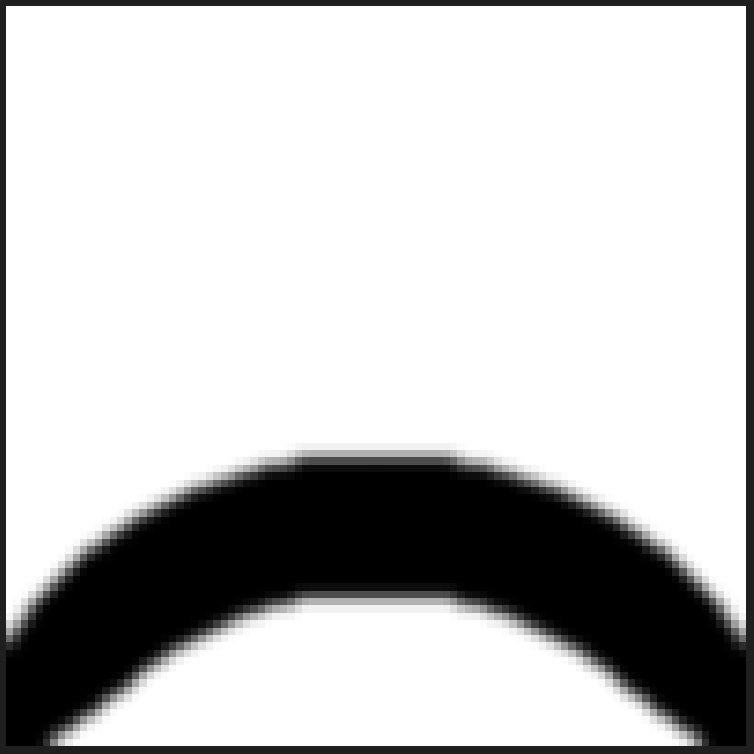}
		\caption{}
		\label{fig:so_press}
	\end{subfigure}
	\quad
	\begin{subfigure}{0.30\textwidth}
		\centering
		\includegraphics[scale=0.20]{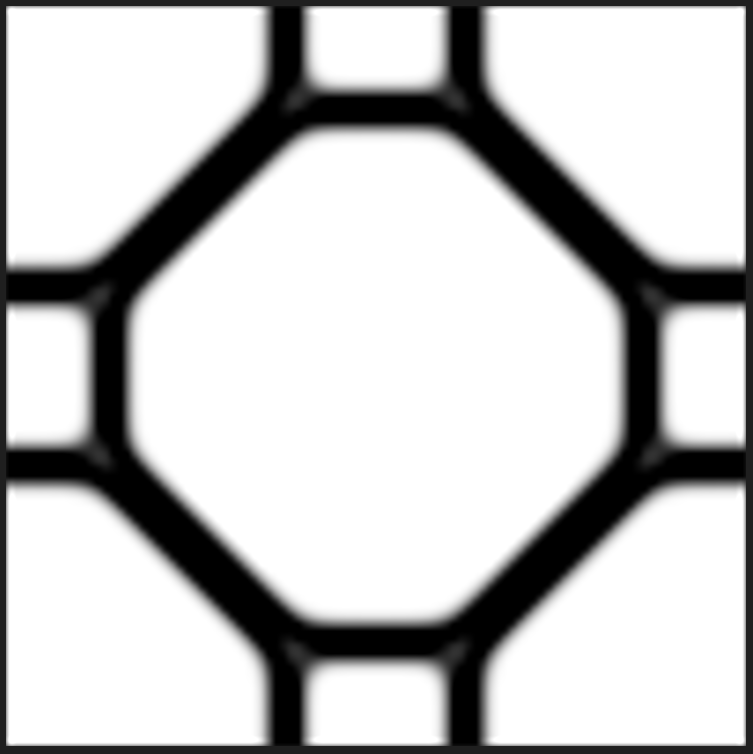}
		\caption{}
		\label{fig:so_mid}
	\end{subfigure}
	\caption{Training data set: Input and target images. (\subref{fig:inputimage}) Input image. Target images: (\subref{fig:so_ss_full}) Mid-load (Fig.~\ref{fig:mid_load}), (\subref{fig:so_x}) Cantilever beam with center load (Fig.~\ref{fig:can_1cent}), (\subref{fig:so_bot}) Cantilever beam a load at free end (Fig.~\ref{fig:can_1end}), (\subref{fig:so_press}) Pressure loadbearing arch structure (Fig.~\ref{fig:archDD})  (\subref{fig:so_mid}) Bulk modulus optimization problem (Fig.~\ref{fig:toxDD})}
	
	\label{fig:Training set}
\end{figure*}
\section{Generation of training data}\label{Sec:GentrainData}
We use a convolutional neural network in the proposed model (Fig.~\ref{fig:CNNarchitech}); hence, both the network's input and output must be images. For generating the output training data (optimized designs, cf.~Fig.~\ref{fig:Training set}),  $\texttt{top88}$~\citep{andreassen2011efficient}, $\texttt{TOPress}$~\citep{kumar2023TOPress} and $\texttt{topX}$~\citep{xia2015design} MATLAB codes are employed for constant load problem (Fig.~\ref{fig:mid_load}, Fig.~\ref{fig:can_1cent} and Fig.~\ref{fig:can_1end}) and for design-dependent load problem (Fig.~\ref{fig:archDD}), respectively. The output images have size $100\times 100$ (i.e., $100\times100$ FEs are used to parameterize the design domains). We vary the volume fraction from 0.01 to 0.95 with an increment of 0.01 to obtain the outputs of the training data. $\{\mt{penal =3,\, rmin =2.4,\,ft=1}\}$,  $\{\mt{penal =3},\, \mt{rmin=2.4},\,\mt{etaf = 0.2},\,\mt{betaf =8}, \linebreak\mt{lst = 1},\,\mt{maxit =100}\}$ and $\{\mt{penal =3,\, rmin =2.4,\,ft=1}\}$ are used as inputs to  $\texttt{top88}$, $\texttt{TOPress}$ and $\texttt{topX}$ MATLAB codes for the respective problems (Fig.~\ref{fig:Deasign domains}). For the input images, we generate black-and-white images wherein the fraction of black pixels equals the given volume fractions (Fig.~\ref{fig:inputimage}). This procedure to generate training data gives 95 input and corresponding output images obtained using the MATLAB codes. If generating data for volume fractions ranging from 0.01 to 0.95 proves challenging, users should generate training images within the feasible volume fraction range.

\section{Python Implementation}\label{Sec:Pyimpl}
This section describes the Python implementation of the proposed CNN architecture. We use the Jupyter Notebook environment. TensorFlow and Keras used to develop and train the model. The code contains five main parts, which are described below.

\subsection{Installing Tensorflow and importing important libraries}
We first install the \ttt{TensorFlow} in our system using the following command on line~3.
\begin{lstlisting}[basicstyle=\scriptsize\ttfamily,breaklines=true,numbers=none,frame=tb]
pip install tensorflow
\end{lstlisting}
\ttt{TensorFlow} is developed by the GoogleBrain team. It facilitates various machine learning tasks, e.g., classification, regression, and clustering, to name a few. \ttt{pip install Libname}, wherein \ttt{Libname} indicates the operation required libraries. Lines~4-10 import other required libraries that come pre-installed in the jupyter environment setup as
\begin{lstlisting}[basicstyle=\scriptsize\ttfamily,breaklines=true,numbers=none,frame=tb]
import os
import numpy as np
import matplotlib.pyplot as plt
import tensorflow as tf 
from matplotlib.image import imread
from tensorflow.keras.models import Sequential
from tensorflow.keras.layers import Conv2D, Conv2DTranspose, MaxPool2D
\end{lstlisting}
\ttt{os} module facilitates operating system-dependent functionality, e.g., file-keeping/calling activities.  \ttt{numpy} is a numerical computing library.  \ttt{conv2D}, \ttt{Conv2DTranspose} and \ttt{MaxPool2D} indicate convolution (Sec.~\ref{Sec:ConOp}),  transpose convolution (Sec.~\ref{Sec:TConOp}) and Maxpooling (Sec.~\ref{Sec:MaxPooling}), respectively.

\subsection{Providing training data file path}
Create \ttt{input\_data} and \ttt{output\_data} folders in the main folder named \ttt{main}.
The training data file is provided using the following piece of codes (line~11-15)
\begin{lstlisting}[basicstyle=\scriptsize\ttfamily,breaklines=true,numbers=none,frame=tb]
Input_train_folder_path ='main\input_data'
Output_train_folder_path ='main\output_data'
Input_train_elements = os.listdir(Input_train_folder_path) 
Output_train_elements = os.listdir(Output_train_folder_path)
\end{lstlisting}
The first two lines (lines~12-13) store the file path of the training data input and output as strings in the respective variables. The following two lines (lines~14-15) save the list of all the elements (training inputs and outputs) in the mentioned file path in the respective variables.
\subsection{Developing the input and output training tensors}
In this section, we develop the input and out training tensors that is supplied to train the model using the following codes (lines~16-28).
\begin{lstlisting}[basicstyle=\scriptsize\ttfamily,breaklines=true,numbers=none,frame=tb]
Input_train = np.zeros((95,100,100,1))
Output_train= np.zeros((95,100,100,1))
for index, Input_train_element in enumerate(Input_train_elements):
element_path = os.path.join(Input_train_folder_path, Input_train_element)
img = imread(element_path)
img = img.reshape((100, 100, 1))
Input_train[index] = img   
for index, Output_train_element in enumerate(Output_train_elements):
element_path = os.path.join(Output_train_folder_path, Output_train_element)
img = imread(element_path)
img = img.reshape((100, 100, 1))
Output_train[index] = img  
\end{lstlisting}
The first two lines (lines~17-18) initialize the input and output training tensors. Both tensors have a size of (95,100,100,1), wherein the first entry, i.e., 95 indicates the number of training examples supplied, (100,\, 100,\,1) indicate pixel size in $x-$, $y-$ and $z-$directions, respectively. The channel size ($z-$direction) is 1 using a gray image. We use a \ttt{for} loop to develop the training input tensor \ttt{Input\_train}. It stores each image's pixel brightness intensity values at their respective position. Likewise, the output tensor \ttt{Output\_train} is developed.

\subsection{Developing the CNN model}
This part of the code develops the proposed CNN architecture (line~29-44). 
\begin{lstlisting}[basicstyle=\scriptsize\ttfamily,breaklines=true,numbers=none,frame=tb]
model = Sequential()
#CONVOLUTIONAL LAYER 1
model.add(Conv2D(filters=128,kernel_size=(2,2),strides=(1,1),padding='same',input_shape=(100,100,1),activation='relu'))
#MAXPOOLING LAYER
model.add(MaxPool2D(pool_size=(2,2),strides=(2,2)))
#CONVOLUTIONAL LAYER 2
model.add(Conv2D(filters= 256,kernel_size=(2,2),strides=(1,1),padding='same',activation='relu'))
#MAXPOOLING LAYER
model.add(MaxPool2D(pool_size=(2,2),strides=(2,2)))
#CONVOLUTIONAL LAYER 3
model.add(Conv2D(filters= 512,kernel_size=(5,5),strides=(1,1),padding='same',activation='relu'))
#MAXPOOLING LAYER
model.add(MaxPool2D(pool_size=(5,5),strides=(5,5)))
## FLATTENING THE ABOVE LAYER
# DENSE LAYER 1
model.add(tf.keras.layers.Flatten())
# DENSE LAYER 2
model.add(tf.keras.layers.Dense(units=8000, activation='relu'))
# DENSE LAYER 3
model.add(tf.keras.layers.Dense(units=12800, activation='relu'))          
# RESHAPING THE ABOVE LAYER
model.add(tf.keras.layers.Reshape(target_shape=(5,5,512)))
#TRANSPOSE CONVOLUTIONAL LAYER 1
model.add(Conv2DTranspose(filters=256,kernel_size=(2,2),strides=(2,2),
activation='relu'))
#TRANSPOSE CONVOLUTIONAL LAYER 2
model.add(Conv2DTranspose(filters=128,kernel_size=(5,5),strides=(5,5),
activation='relu'))
#TRANSPOSE CONVOLUTIONAL LAYER 3
model.add(Conv2DTranspose(filters=1,kernel_size=(2,2),strides=(2,2),
activation='relu'))
#COMPILING THE MODEL
model.compile(optimizer='adam', loss='mean_squared_error')
\end{lstlisting}
Each layer of the proposed network contains single input and output tensors; thus, the sequential model is used herein, which is simple. The proposed network can be defined step-by-step; that is, the network can be built by adding one layer at a time. Sequential is defined as (line~30)
\begin{lstlisting}[basicstyle=\scriptsize\ttfamily,breaklines=true,numbers=none,frame=tb]
model = Sequential() 
\end{lstlisting}
To incorporate different layers in the sequential model,  \ttt{model.add()} function is utilized. We start with the first convolution layer; for that, \ttt{Conv2D} function is employed (line~31). The key parameters of the \ttt{Conv2D} function are:
	\begin{itemize}
	\item \underline{filters}:  It is the number of kernels or channels in the convolutional layer. Each filter detects different features in the input.
      \item \underline{kernel size} : Size of the convolutional kernels. It can be a single integer or a tuple (height, width) for square and rectangular kernels.
       \item \underline{strides}: The step size used by the convolutional kernel as it moves across the input. It is a tuple (stride vertical, stride horizontal).
       \item \underline{padding} : This parameter determines the padding strategy.
       \item \underline{activation}: An activation function applied to convolution's output.
         \item \underline{input shape}:  It specifies the shape of the input data. 
	\end{itemize}
\ttt{maxPool2D} function (line~32) is used for the maxpooling purposes (Sec.~\ref{Sec:MaxPooling}) with \ttt{pool\_size} of (2,\,2). Likewise, convolutional layer 2 (line~33) and layer 3 (line~35) are defined, which are followed by  \ttt{maxPool2D} function (line~34 and line~36). The output of the last max-pooling layer is flattened to form the first dense layer with 12800 neurons using \ttt{tf.keras.layers.Flatten()} function (line~37). After that, the adaptive layer (line~38) and third layers are added using \ttt{tf.keras.layers.Dense} function (line~39). Note, one comments out/remove the adaptive layer while using the \text{base} architecture. \ttt{units} and \ttt{activation} are the key parameters, wherein the former specifies the number of neurons or nodes in the dense layer. On line~40, the output of the last layer is reshaped using function \ttt{tf.keras.layers.Reshape}. Lines~41-43 add three transpose convolutional layers using \ttt{Conv2DTranspose} function. The function takes same parameter as \ttt{Conv2D} function. On line~44, the model is compiled using \ttt{model.compile} function that also specifies the optimizer and loss functions used. \ttt{optimizer} indicates the optimization algorithm used during training. We use `\ttt{adam},' which stands for Adaptive Moment Estimation during training (line~44). It adapts the learning rates of each parameter based on their past gradients, providing a balance between efficiency and simplicity. The parameter `loss' specifies the loss function that the model uses during training (line~44). The `mean squared error' is used, as the goal is to minimize the squared difference between the predicted and actual values (line~44). Next, training is performed.
\begin{table*}[ht]
	\caption{Optimized design for mid-load problem (Fig.~\ref{fig:Deasign domains}). $V_\text{err}$ and $Obj_\text{err}$ are the volume fraction and objective errors between CNN results and corresponding target outputs generated by MATLAB code, \ttt{top88}.}
	\label{Table:CNNresults1}
	\centering
	\begin{tabular}{|c|c|c|c|c|c|}
		\hline
		\textbf{$V_f$}& \textbf{Input} & \textbf{CNN results} & \textbf{Target} & \textbf{ $V_\text{err}$ (\%)} & \textbf{$Obj_\text{err}$ (\%)}\\
		\hline
		0.05 & \includegraphics[width=2cm]{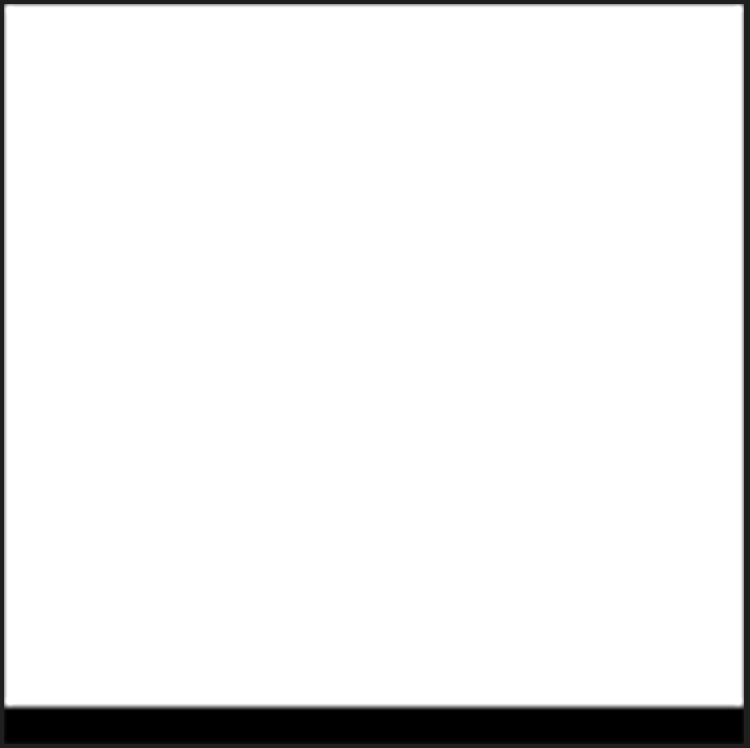} & \includegraphics[width=2cm]{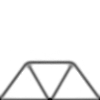} & \includegraphics[width=2cm]{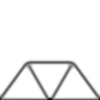} & 5.4 & 1.65 \\
		\hline
		0.15 & \includegraphics[width=2cm]{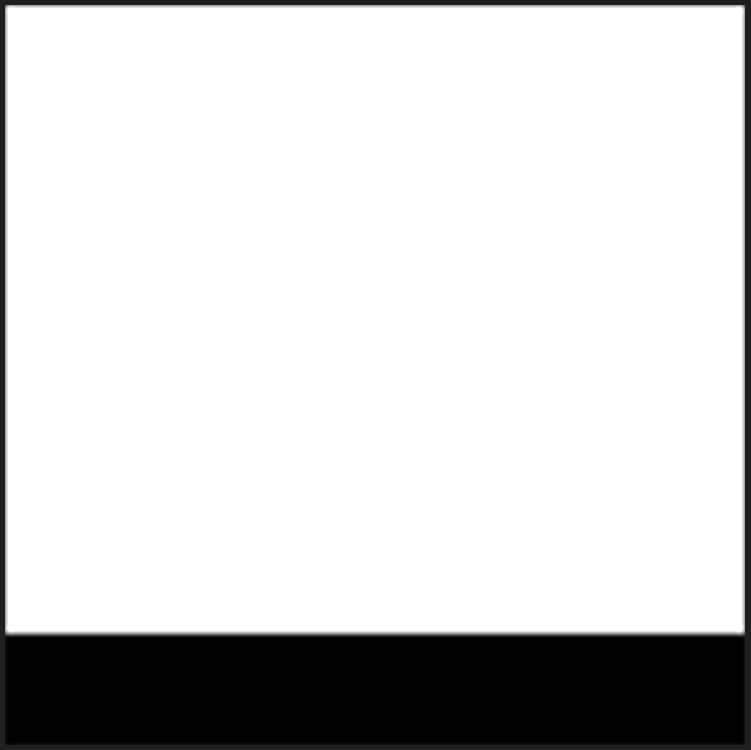} & \includegraphics[width=2cm]{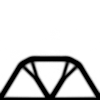} & \includegraphics[width=2cm]{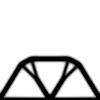} & 0.53 & 2.66 \\
		\hline
		0.20 & \includegraphics[width=2cm]{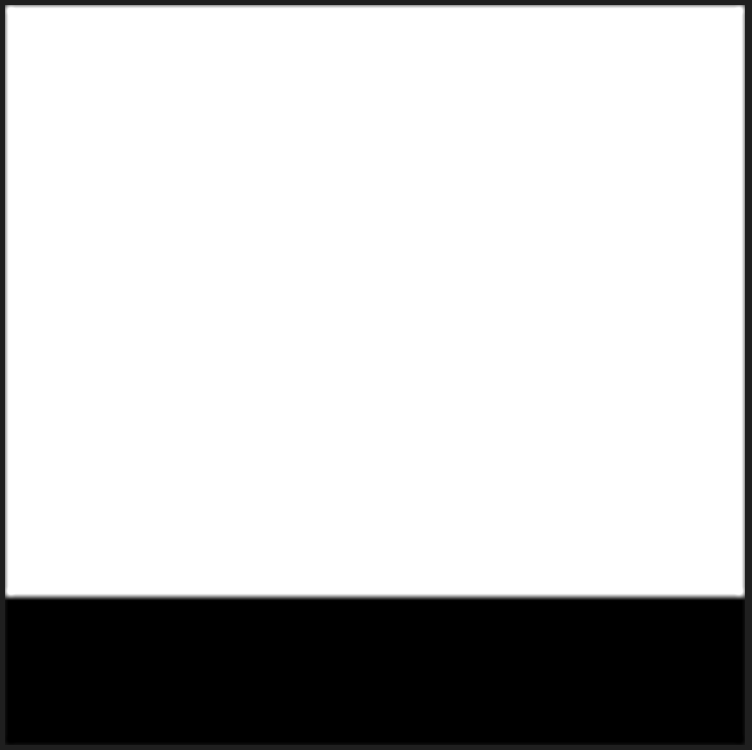} & \includegraphics[width=2cm]{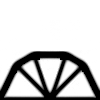} & \includegraphics[width=2cm]{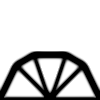} & 0.20 & 0.041\\
		\hline
		0.25 & \includegraphics[width=2cm]{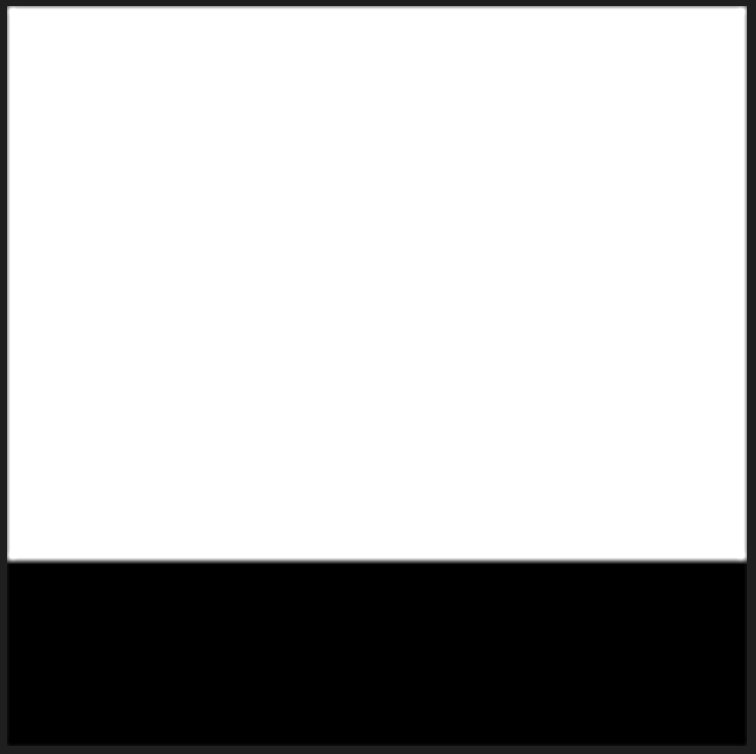} & \includegraphics[width=2cm]{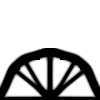} & \includegraphics[width=2cm]{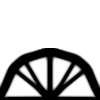} & 0.48 & 0.51\\
		\hline
		0.35 & \includegraphics[width=2cm]{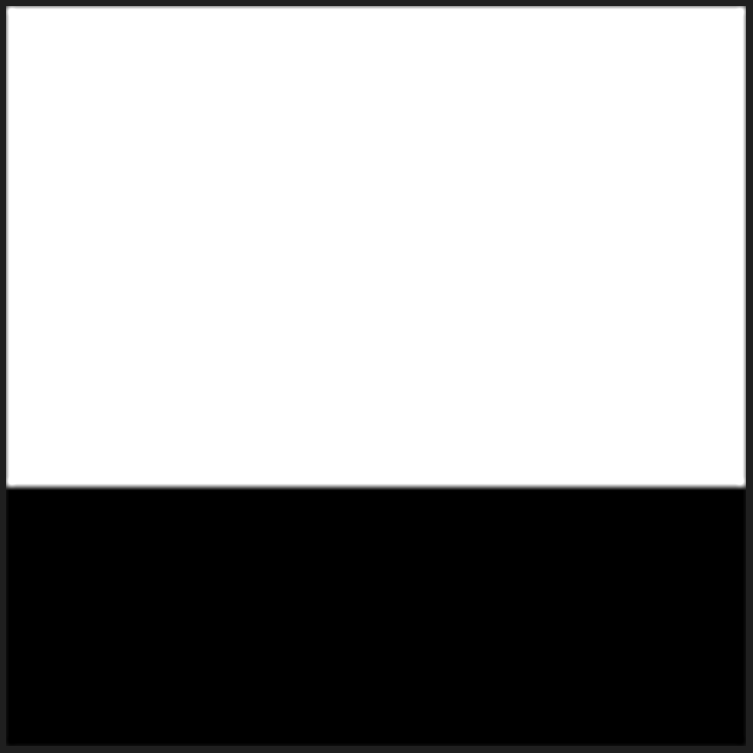} & \includegraphics[width=2cm]{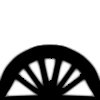} & \includegraphics[width=2cm]{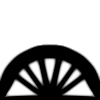} & 0.057 & 0.144\\
		\hline
		0.50 & \includegraphics[width=2cm]{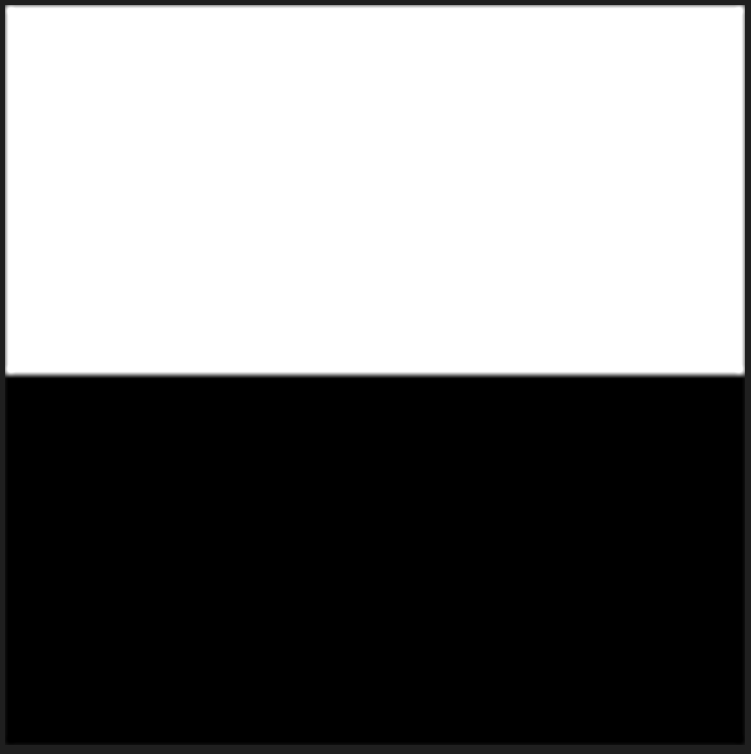} & \includegraphics[width=2cm]{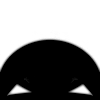} & \includegraphics[width=2cm]{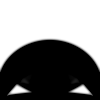} & 0.10 & 0.075\\
		\hline
		0.75 & \includegraphics[width=2cm]{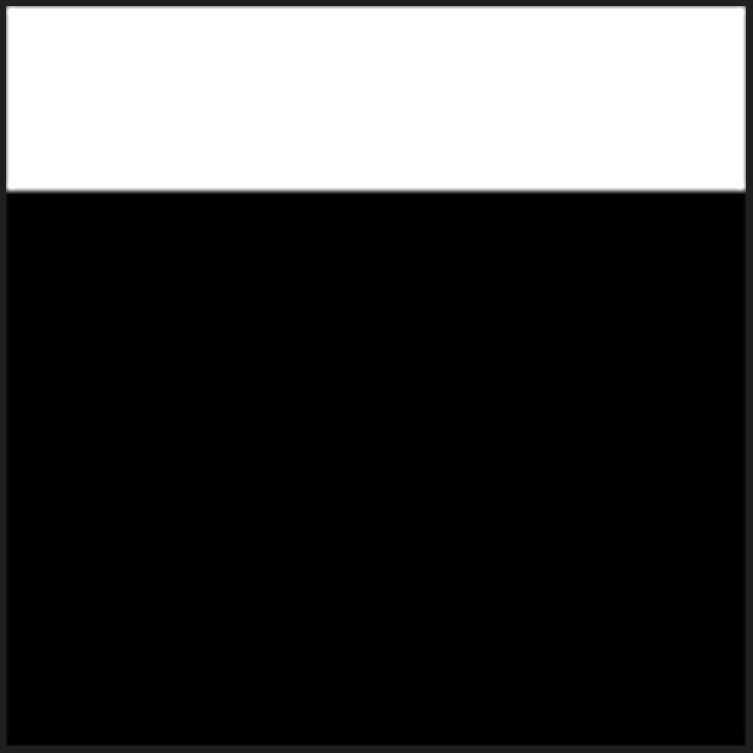} & \includegraphics[width=2cm]{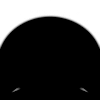} & \includegraphics[width=2cm]{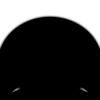} & 0.053 & 0.096\\
		\hline
	\end{tabular}
\end{table*}

\begin{figure}[h!]
	\centering
	\begin{subfigure}{0.20\textwidth}
		\centering
		\includegraphics[scale = 0.40]{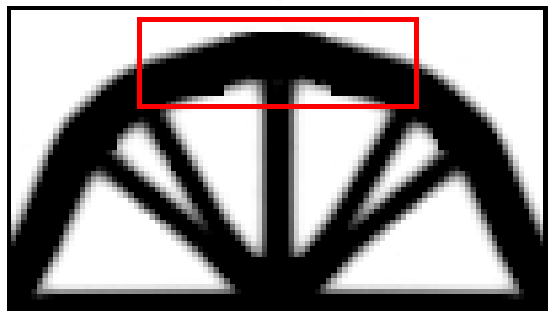}
		\caption{CNN output}
		\label{fig:ss_pred_25_error}
	\end{subfigure}
	\hfill
	\begin{subfigure}{0.20\textwidth}
		\centering
		\includegraphics[scale = 0.40]{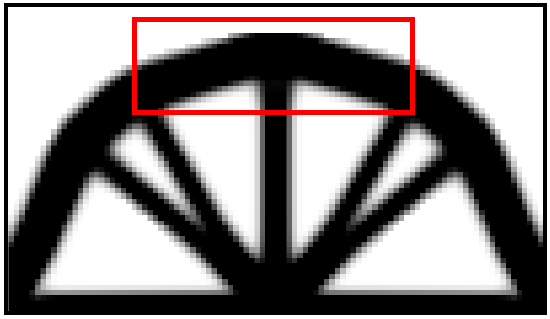}
		\caption{Target design}
		\label{fig:ss_targ_25}
	\end{subfigure}
	\caption{Results comparison for volume fraction 0.25}
	\label{fig:comparison_ss}
\end{figure}
\subsection{Training command}\label{Sec:TrainComm}
In this part of the code, the training of the model is performed (line~) using the following piece of codes.
\begin{lstlisting}[basicstyle=\scriptsize\ttfamily,breaklines=true,numbers=none,frame=tb]
model.summary()
model.fit(Input_train,Output_train,epochs = 2000)
\end{lstlisting}
\ttt{model.summary} (line~46) provides the architecture's summary, including information about the layers, the number of parameters in each layer, and the total number of parameters in the model. \ttt{model.fit} (line~47) instructs to train the model on a data set that has been provided (\ttt{Input\_data}, \ttt{Output\_data}). ``\ttt{epochs} = 2000" (line~47) indicates that the model should undergo 2000 training iterations on the provided data. 

\section{Results and discussion}\label{Sec:resndis}
\label{sec:resndis}
In this section, we generate optimized designs for problems mentioned in Fig.~\ref{fig:Deasign domains} to demonstrate the versatility of the proposed CNN. 

When applying the proposed architecture (Fig.~\ref{fig:CNNarchitech}) to automate various optimization problems, it is observed that in the majority of cases, the base architecture, post training, generated accurate optimized designs that closely resembled those produced by the problem specific open-source MATLAB codes. In problems where the outputs generated by the base architecture fail to meet the required standards, incorporating the adaptive layer may enhance quality of the optimized designs produced by the CNN. This is because addition of the adaptive layer significantly increases the number of learnable parameters in the network which increases capacity of the model to map complex functions to a large extent

We determine volume fraction and objective errors between the results provided by the CNN and MATLAB codes to note the closeness between the results. As the network gives the output image in grayscale, we get the $\ttt{xphys}_\text{CNN}$ vector from it. $V_\text{CNN} = \ttt{mean}(\ttt{xphys}_\text{CNN})\times nel$ is determine; thus, $V_\texttt{err} = \frac{V^*-V_\text{CNN}}{V_f} \times 100 \%$. $nel$ is the number of FEs utilized. Likewise, using  $\ttt{xphys}_\text{CNN}$ in the employed codes (\ttt{top88},\,\ttt{TOPress}, and,\,\ttt{topX}), determine $C_\text{CNN}$ and  $C_\texttt{err} = \frac{C_\text{opt} - C_\text{CNN}}{C_\text{opt}} \times 100 \%$, where $C_\text{opt}$ is the objective value directly obtained from the MATLAB codes for the same volume fractions.

\begin{table*}[h!]
	\caption{Optimized cantilever beam with a load at free end (Fig.~\ref{fig:Deasign domains}). $V_\text{err}$ and $Obj_\text{err}$ are the volume fraction and objective errors between CNN results and corresponding target outputs generated by MATLAB code, \ttt{top88}.}
	\label{Table:CNNresults2}
	\centering
	\begin{tabular}{|c|c|c|c|c|c|}
		\hline
		\textbf{$V_f$}& \textbf{Input} & \textbf{CNN results} & \textbf{Target} & \textbf{\% $V_{err}$} & \textbf{\% $Obj_{err}$}\\
		\hline
		0.10 & \includegraphics[width=2cm]{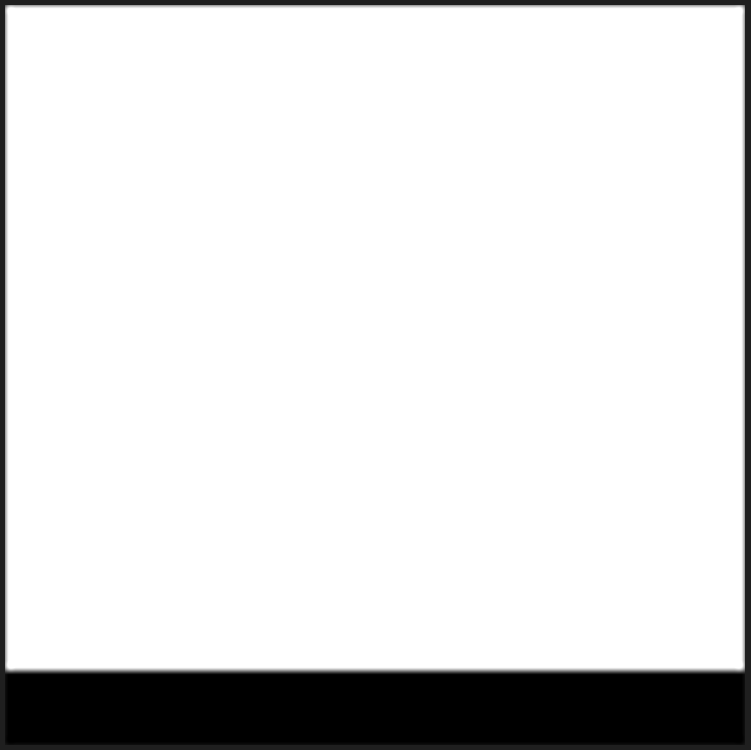} & \includegraphics[width=2cm]{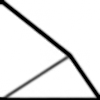} & \includegraphics[width=2cm]{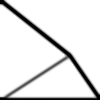} & 1.2 & 0.20 \\
		\hline
		0.25 & \includegraphics[width=2cm]{vf_25_in_final.png} & \includegraphics[width=2cm]{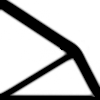} & \includegraphics[width=2cm]{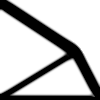} & 0.36 & 0.50 \\
		\hline
		0.40 & \includegraphics[width=2cm]{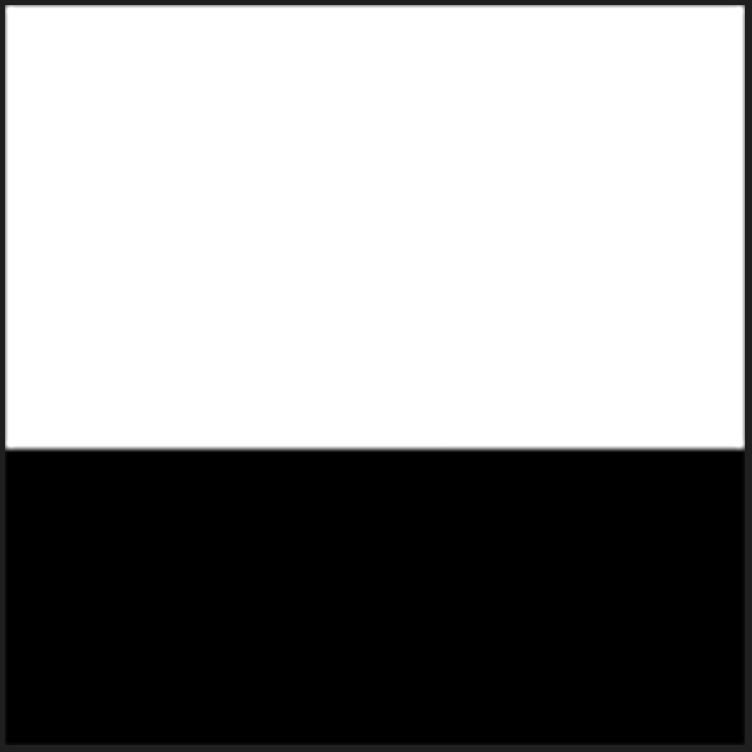} & \includegraphics[width=2cm]{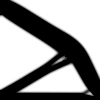} & \includegraphics[width=2cm]{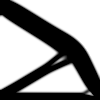} & 0.23 & 0.29 \\
		\hline
		0.50 & \includegraphics[width=2cm]{vf_50_in_final.png} & \includegraphics[width=2cm]{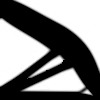} & \includegraphics[width=2cm]{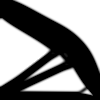} & 0.22 & 0.14 \\
		\hline
		0.60 & \includegraphics[width=2cm]{vf_60_in.png} & \includegraphics[width=2cm]{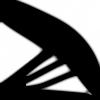} & \includegraphics[width=2cm]{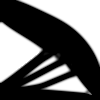} & 0.40 & 0.24 \\
		\hline
		0.70 & \includegraphics[width=2cm]{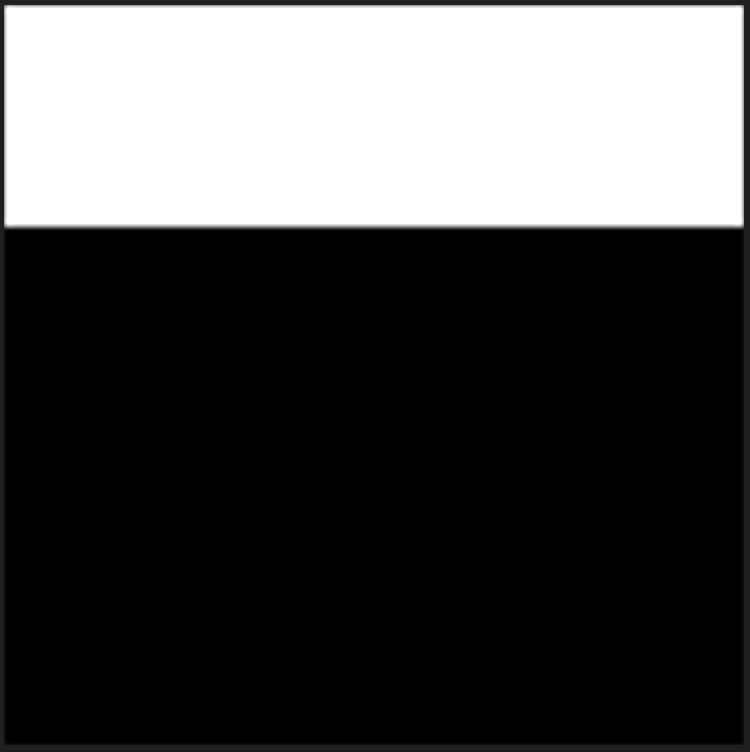} & \includegraphics[width=2cm]{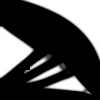} & \includegraphics[width=2cm]{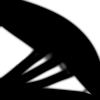} & 0.19 & 0.063 \\
		\hline
		0.85 & \includegraphics[width=2cm]{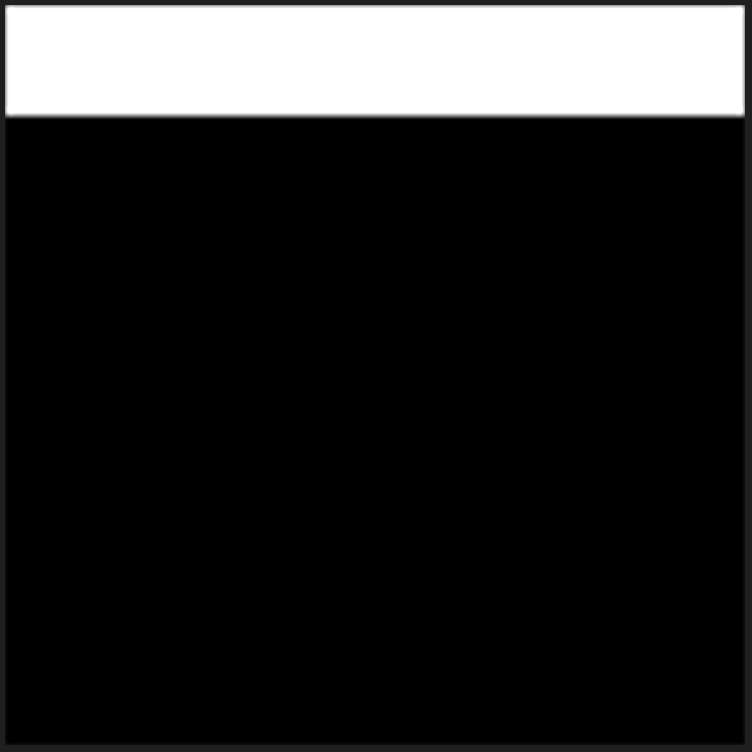} & \includegraphics[width=2cm]{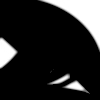} & \includegraphics[width=2cm]{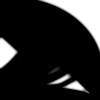} & 0.11 & 0.025 \\
		\hline
	\end{tabular}
\end{table*}
  \begin{table*}[t]
  	\caption{\PKnew{Optimized cantilever beam with a load at free end (Fig.~\ref{fig:Deasign domains}). Input volume fractions are not available in the training datasets.  $V_\text{err}$ and $Obj_\text{err}$ are the volume fraction and objective errors between CNN results and corresponding target outputs generated by MATLAB code, \texttt{top88}.}}
  	\label{Table:CNNresults3c}
  	\centering
  	\begin{tabular}{|c|c|c|c|c|c|}
  		\hline
  		\textbf{$V_f$}& \textbf{Input} & \textbf{CNN results} & \textbf{Target} & \textbf{\% $V_{err}$} & \textbf{\% $Obj_{err}$}\\
  		\hline
  		0.175 & \includegraphics[width=2cm]{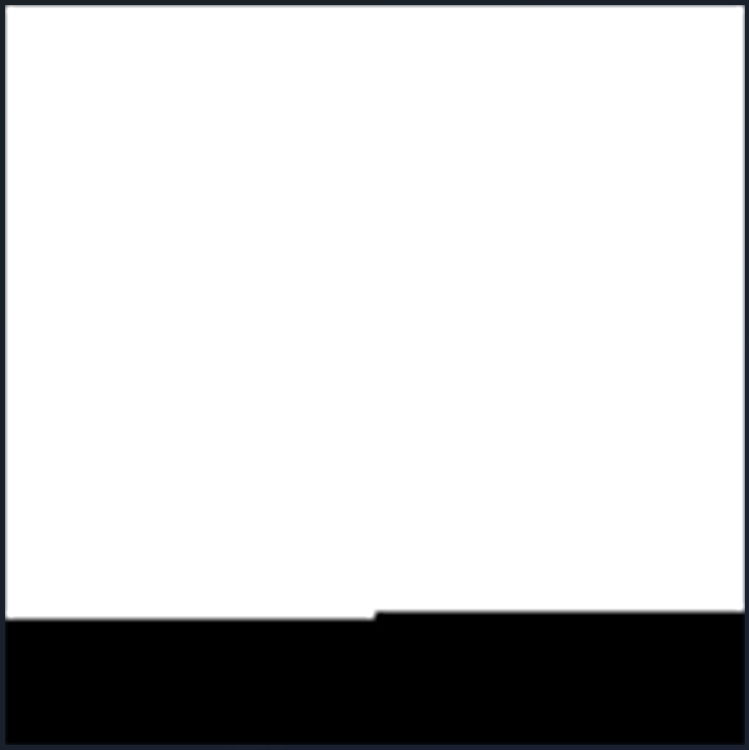} & \includegraphics[width=2cm]{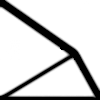} & \includegraphics[width=2cm]{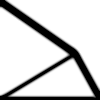} & 2.17 & 0.49 \\
  		\hline
  		0.366 & \includegraphics[width=2cm]{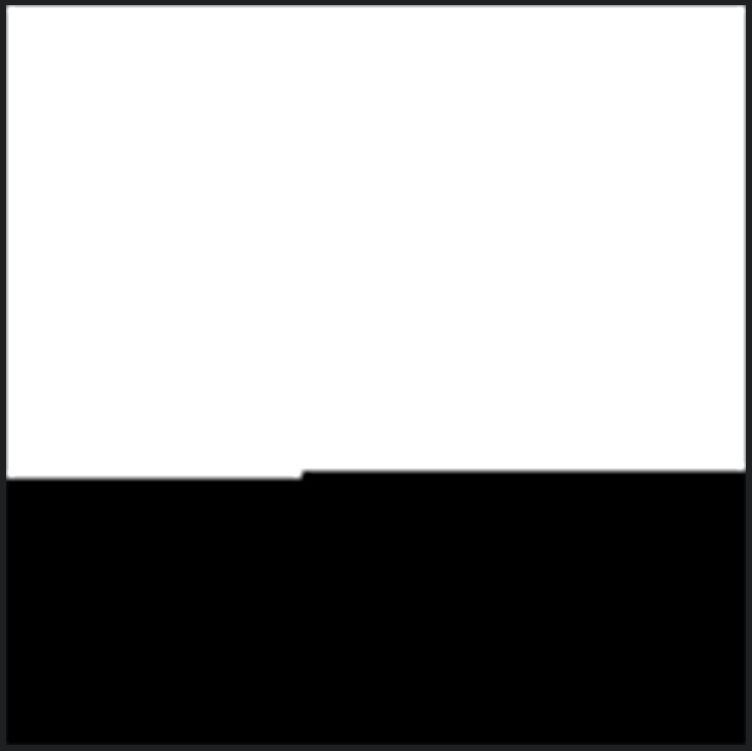} & \includegraphics[width=2cm]{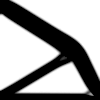} & \includegraphics[width=2cm]{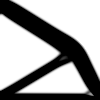} & 0.79 & 0.17 \\
  		\hline
  		0.447 & \includegraphics[width=2cm]{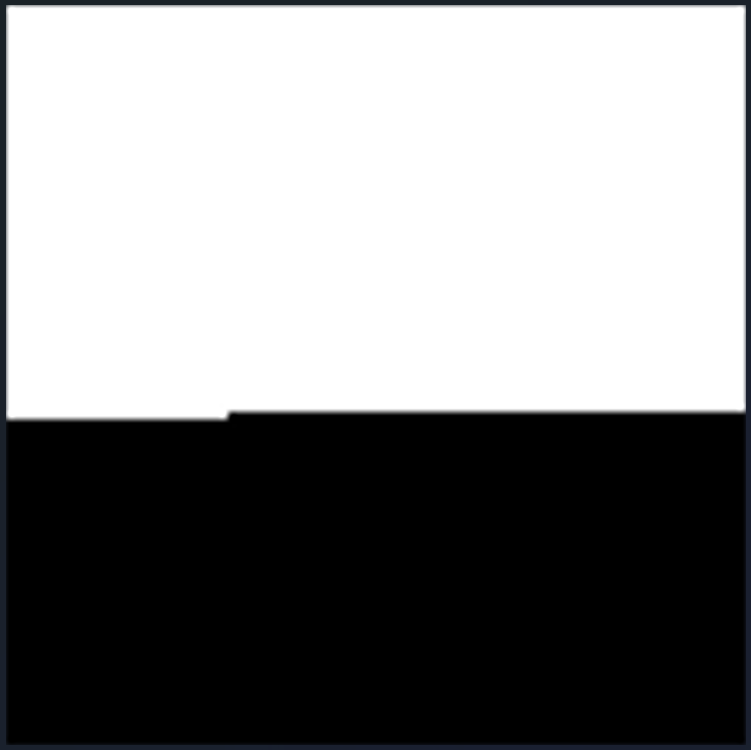} & \includegraphics[width=2cm]{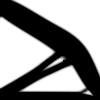} & \includegraphics[width=2cm]{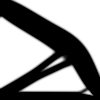} & 0.67 & 0.098 \\
  		\hline
  		0.558 & \includegraphics[width=2cm]{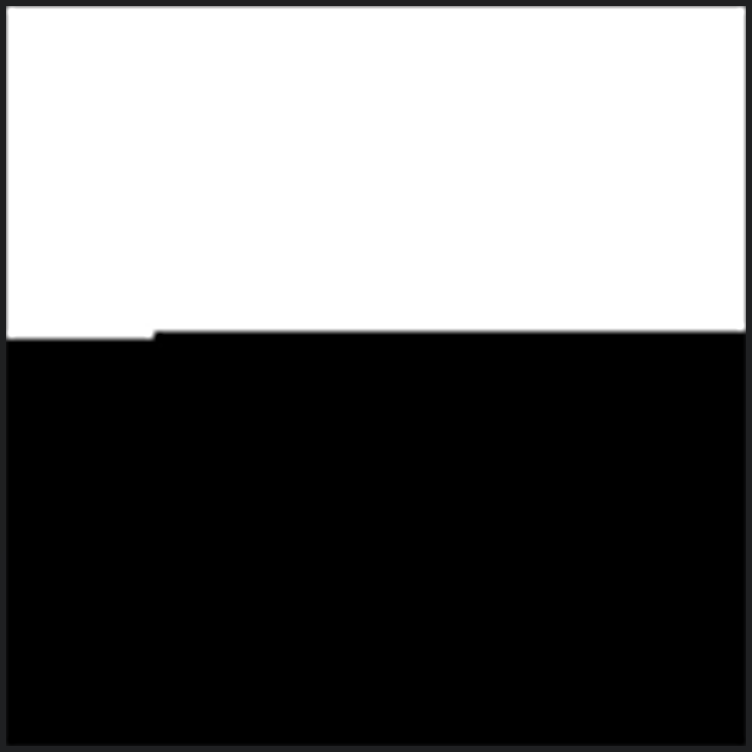} & \includegraphics[width=2cm]{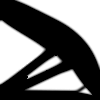} & \includegraphics[width=2cm]{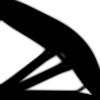} & 0.197 & 0.25 \\
  		\hline
  		0.721 & \includegraphics[width=2cm]{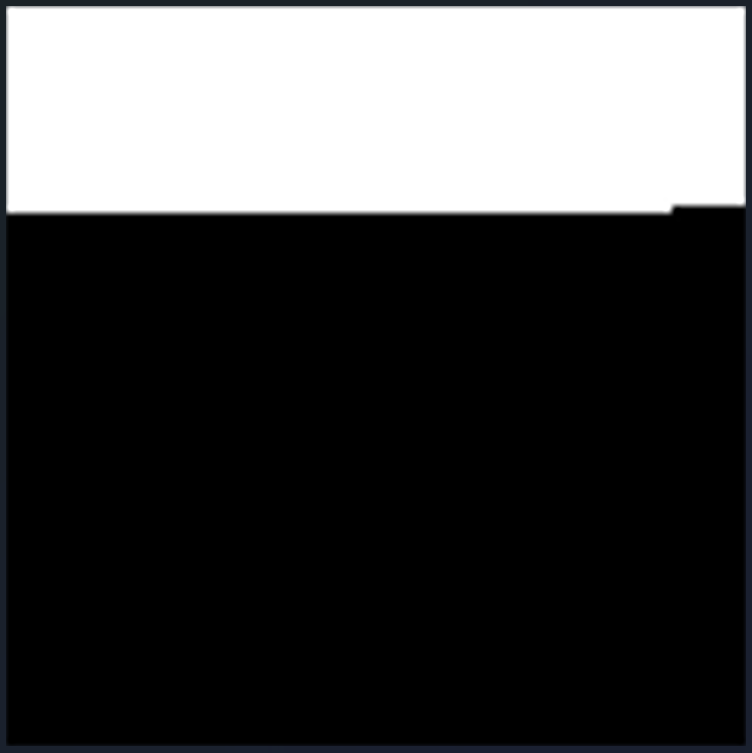} & \includegraphics[width=2cm]{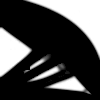} & \includegraphics[width=2cm]{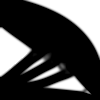} & 0.11 & 0.097 \\
  		\hline
  	\end{tabular}
  \end{table*}
  \begin{table*}[t]
  	\caption{\PKnew{Optimized cantilever beam with a load at free end (Fig.~\ref{fig:Deasign domains}) with low training datasets. Input volume fractions are not available in the training datasets.  $V_\text{err}$ and $Obj_\text{err}$ are the volume fraction and objective errors between CNN results and corresponding target outputs generated by MATLAB code, \texttt{top88}.}}
  	\label{Table:CNNresults3d}
  	\centering
  	\begin{tabular}{|c|c|c|c|c|c|}
  		\hline
  		\textbf{$V_f$}& \textbf{Input} & \textbf{CNN results} & \textbf{Target} & \textbf{\% $V_{err}$} & \textbf{\% $Obj_{err}$}\\
  		0.15 & \includegraphics[width=2cm]{vf_15_in.png} & \includegraphics[width=2cm]{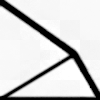} & \includegraphics[width=2cm]{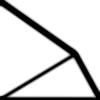} & 14.4 & 2.36 \\
  		\hline
  		0.25 & \includegraphics[width=2cm]{vf_25_in_final.png} & \includegraphics[width=2cm]{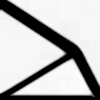} & \includegraphics[width=2cm]{topbot_targ_vf_25_out.png} & 11.2 & 2.03 \\
  		\hline
  		0.40 & \includegraphics[width=2cm]{vf_40_in.png} & \includegraphics[width=2cm]{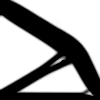} & \includegraphics[width=2cm]{topbot_targ_vf_40_out.png} & 0.85 & 1.22 \\
  		\hline
  		0.60 & \includegraphics[width=2cm]{vf_60_in.png} & \includegraphics[width=2cm]{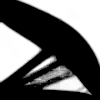} & \includegraphics[width=2cm]{topbot_targ_vf_60_out.png} & 2.2 & 0.07 \\
  		\hline
  		0.70 & \includegraphics[width=2cm]{vf_70_in.png} & \includegraphics[width=2cm]{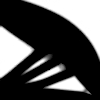} & \includegraphics[width=2cm]{topbot_targ_vf_70_out.png} & 1.31 & 0.41 \\
  		\hline
  		0.85 & \includegraphics[width=2cm]{vf_85_in.png} & \includegraphics[width=2cm]{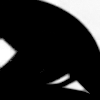} & \includegraphics[width=2cm]{topbot_targ_vf_85_out.png} & 0.39 & 0.13 \\
  		\hline
  	\end{tabular}
  \end{table*}

\subsection{Results with base architecture}\label{Sec:ResWBarch}
The Python code, \texttt{PyTOaCNN} for the proposed CNN model is provided in Appendix~\ref{Sec:PyTOaCNN}. The \textit{base} architecture part is obtained by commenting line~38 (i.e., removing line~38). After that, the training procedure is performed as mentioned in Sec.~\ref{Sec:TrainComm}.

Herein, the derived base architecture is used to obtain the optimized solutions for the mid-load problem (Fig.~\ref{fig:mid_load}),  cantilever beams with end constant load (Fig.~\ref{fig:can_1end}), design-dependent pressure loadbearing arch structure (Fig.~\ref{fig:archDD}), and material bulk modulus optimization problems (Fig.~\ref{fig:toxDD}). The results are depicted in Tables~\ref{Table:CNNresults1}-\ref{Table:CNNresults_topX}. The first and second columns give the utilized volume fraction and input images, respectively. The third and fourth columns provide results obtained by the proposed CNN model and by the MATLAB codes,  respectively. The fifth and sixth columns furnish the volume fraction and objective errors, respectively.  
\begin{figure}[hbt]
	\centering
	\begin{subfigure}{0.20\textwidth}
		\centering
		\includegraphics[scale = 0.8]{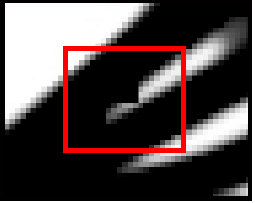}
		\caption{CNN output}
		\label{fig:bot_pred_70_error}
	\end{subfigure}
	\hfill
	\begin{subfigure}{0.20\textwidth}
		\centering
		\includegraphics[scale = 0.8]{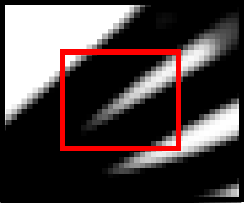}
		\caption{Target design}
		\label{fig:bot_targ70}
	\end{subfigure}
	\caption{Optimized designs comparison for volume fraction 0.70}
	\label{fig:comparison_bot}
\end{figure}
\subsubsection{Mid-load problem}
The mid-load problem is solved first. The design domain, boundary conditions, and the applied load are shown in Fig.~\ref{fig:mid_load}. We generate 95 training samples by varying the volume fraction from 0.01 to 0.95 with an increment of 0.01 using \ttt{top88} MATLAB code (Sec.~\ref{Sec:GentrainData}). We train the base architecture using the generated data for 2000 epochs per Sec.~\ref{Sec:TrainComm}. 

Results obtained with different volume fractions are shown in Table~\ref{Table:CNNresults1}, obtained in a fraction of a second. One notes that the proposed model can yield accurate, optimized designs with low volume fraction and objective function value errors~(Table ~\ref{Table:CNNresults1}). For the volume fraction 0.20, the $Obj_\text{err}$ is minimum, whereas it is higher for volume fraction 0.15. Results obtained using the proposed CNN model closely (exactly) resemble those obtained using the MATLAB code. Therefore, the model successfully captures the most complex and intricate patterns in the optimized designs. A closer look reveals that the output results have, by and large, the same features as those of the MATLAB code but may not precisely resemble them. However, such negligible deviations have insignificant effects on the performance. For example, the result shown in Fig. ~\ref{fig:comparison_ss} for volume fraction 0.25. Corresponding $V_\text{err}$ and $Obj_\text{err}$ are 0.48 and 0.51, respectively, close to 0.5\%, i.e., negligible. Note that the number of training data used is significantly less than the previous efforts in this directions~\citep{ramu2022survey} still, the proposed model successfully provides the optimized designs with marginal errors.

\begin{table*}
	\caption{Pressure loadbearing arch structure (Fig.~\ref{fig:Deasign domains}). $V_\text{err}$ and $Obj_\text{err}$ are the volume fraction and objective errors between CNN results and corresponding target outputs generated by MATLAB code, \texttt{TOPress}.}
	\label{Table:CNNresults_topress}
	\centering
	\begin{tabular}{|c|c|c|c|c|c|}
		\hline
		\textbf{$V_f$}& \textbf{Input} & \textbf{CNN results} & \textbf{Target} & \textbf{\% $V_{err}$} & \textbf{\% $Obj_{err}$} \\
		\hline
		0.05 & \includegraphics[width=2cm]{vf_in_5.PNG} & \includegraphics[width=2cm]{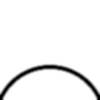} & \includegraphics[width=2cm]{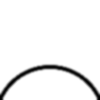} & 4.20 & 4.90 \\
		\hline
		0.15 & \includegraphics[width=2cm]{vf_15_in.png} & \includegraphics[width=2cm]{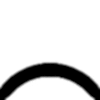} & \includegraphics[width=2cm]{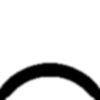} & 1.86 & 0.46 \\
		\hline
		0.25 & \includegraphics[width=2cm]{vf_25_in_final.png} & \includegraphics[width=2cm]{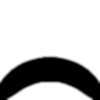} & \includegraphics[width=2cm]{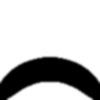} & 0.96 & 0.048 \\
		\hline
		0.35 & \includegraphics[width=2cm]{vf_35_in.png} & \includegraphics[width=2cm]{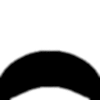} & \includegraphics[width=2cm]{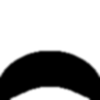} & 0.45 & 0.039 \\
		\hline
		0.45 & \includegraphics[width=2cm]{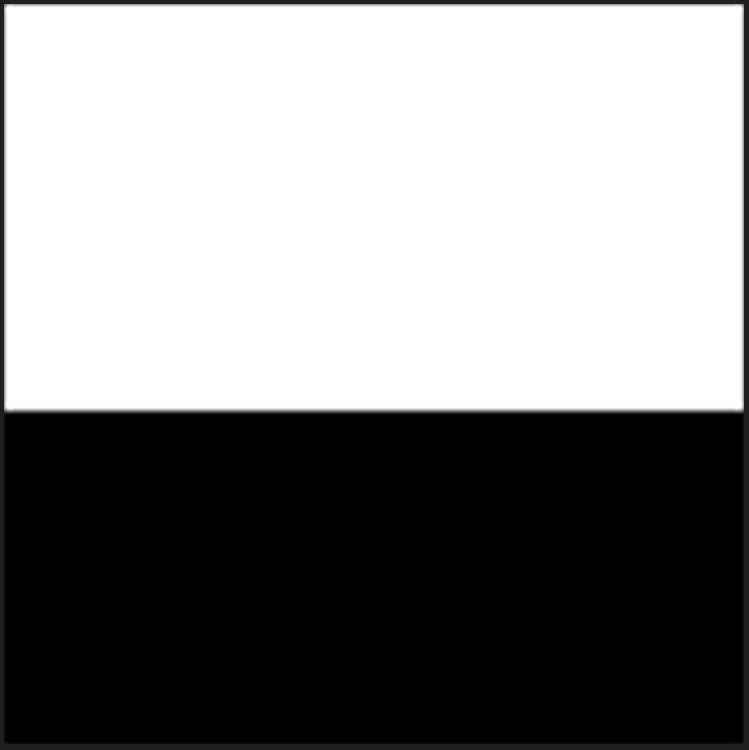} & \includegraphics[width=2cm]{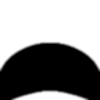} & \includegraphics[width=2cm]{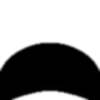} & 0.22 & 0.065 \\
		\hline
		0.55 & \includegraphics[width=2cm]{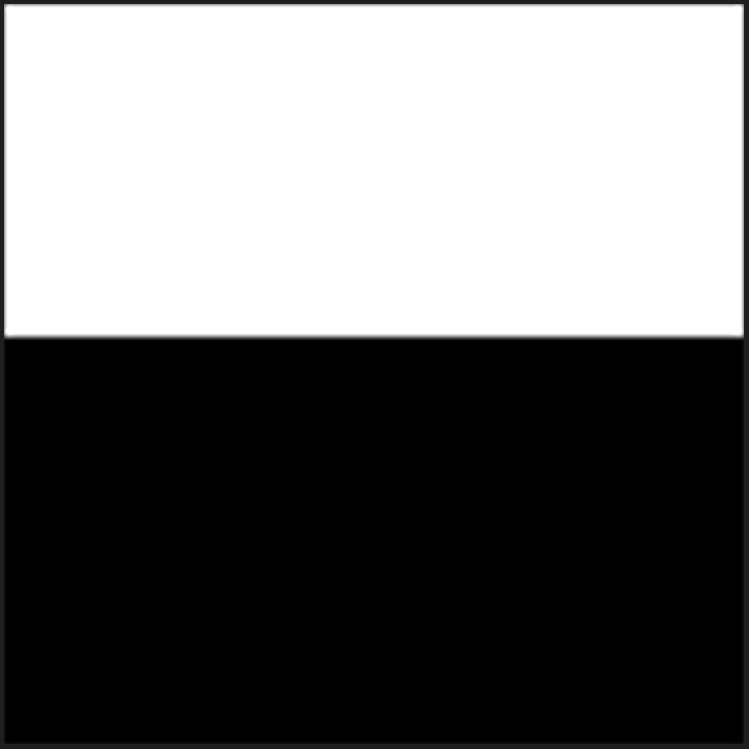} & \includegraphics[width=2cm]{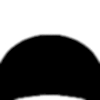} & \includegraphics[width=2cm]{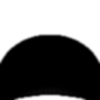} & 0.11 & 0.87 \\
		\hline
		0.75 & \includegraphics[width=2cm]{vf_75_in_final.png} & \includegraphics[width=2cm]{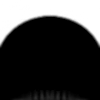} & \includegraphics[width=2cm]{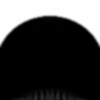} & 0.08 & 0.29 \\
		\hline
	\end{tabular}
\end{table*}
\begin{table*}[h!]
	\caption{Bulk Modulus optimization problem (Fig.~\ref{fig:Deasign domains}). $V_\text{err}$ and $Obj_\text{err}$ are the volume fraction and objective errors between CNN results and corresponding target outputs generated by MATLAB code, \texttt{TopX}.}
	\label{Table:CNNresults_topX}
	\centering
	\begin{tabular}{|c|c|c|c|c|c|c|}
		\hline
		\textbf{$V_f$}& \textbf{Input} & \textbf{CNN results} & \textbf{Target} & \textbf{\% $V_{err}$} & \textbf{\% $Obj_{err}$} \\
		\hline
		0.10 & \includegraphics[width=2cm]{vf_10_in.PNG} & \includegraphics[width=2cm]{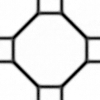} & \includegraphics[width=2cm]{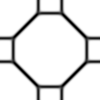} & 6.4 & 3 \\
		\hline
		0.20 & \includegraphics[width=2cm]{vf_20_in.png} & \includegraphics[width=2cm]{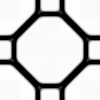} & \includegraphics[width=2cm]{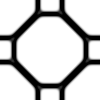} & 5.15 & 0.43\\
		\hline
		0.25 & \includegraphics[width=2cm]{vf_25_in_final.png} & \includegraphics[width=2cm]{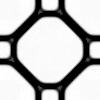} & \includegraphics[width=2cm]{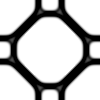} & 4 & 0.12 \\
		\hline
		0.30 & \includegraphics[width=2cm]{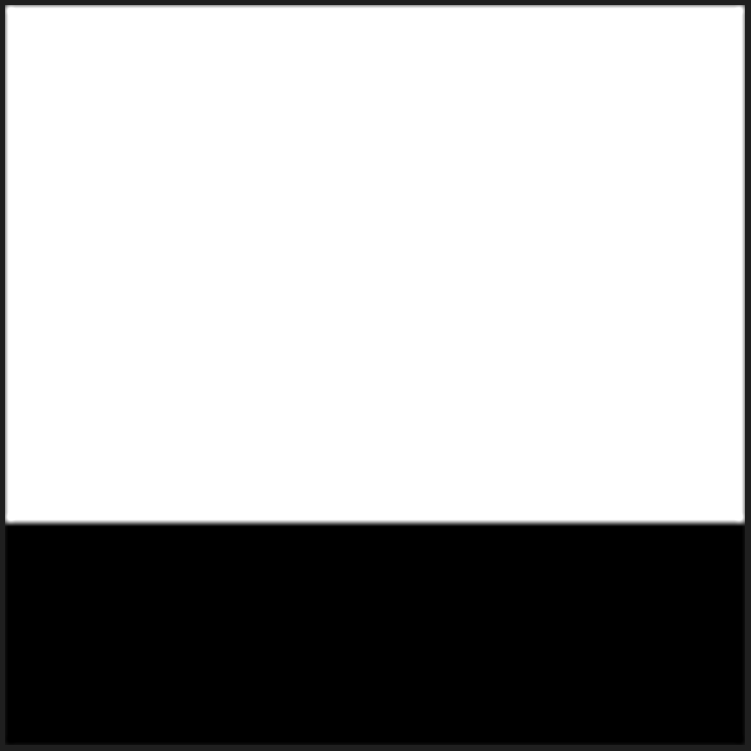} & \includegraphics[width=2cm]{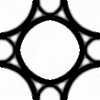} & \includegraphics[width=2cm]{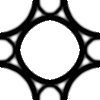} & 1.67 & 0.61 \\
		\hline
		0.35 & \includegraphics[width=2cm]{vf_35_in.png} & \includegraphics[width=2cm]{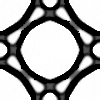} & \includegraphics[width=2cm]{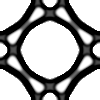} & 1.31 & 1.38 \\
		\hline
		0.40 & \includegraphics[width=2cm]{vf_40_in.PNG} & \includegraphics[width=2cm]{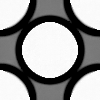} & \includegraphics[width=2cm]{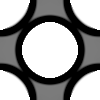} & 2.325 & 0.112 \\
		\hline
		0.50 & \includegraphics[width=2cm]{vf_50_in_final.png} & \includegraphics[width=2cm]{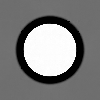} & \includegraphics[width=2cm]{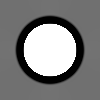} & 1.10 & 1.86 \\
		\hline
	\end{tabular}
\end{table*}
\begin{table*}[h!]
	\caption{Optimized cantilever beam with center load with different $n$ in the adaptive layer (Fig.~\ref{fig:Deasign domains}). $V_\text{err}$ and $Obj_\text{err}$ indicate the volume and objective error between the results obtained by the proposed CNN and the target output.}
	\label{Table:CNNresults_mid_adap}
	\resizebox{\textwidth}{!}{
		\begin{tabular}{|l|l|cccccc|l|}
			\hline
			\textbf{$V_f$} & \textbf{Input} & \multicolumn{6}{c|}{\textbf{CNN results}} & \textbf{Target} \\
			\hline
			\multirow{7}{*}{0.05} & \multirow{7}{*}{\includegraphics[scale=0.075]{vf_in_5.PNG}} & \multicolumn{6}{c|}{\textbf{\# neurons in adaptive layer ($n$)}} & \multirow{7}{*}{\includegraphics[scale=0.5]{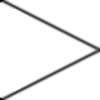}} \\
			\cline{3-8}
			& & \textbf{0} & \textbf{1000} & \textbf{2000} & \textbf{4000} & \textbf{8000} & \textbf{12000} & \\
			\cline{3-8}
			& & \multicolumn{1}{c|}{\includegraphics[scale=0.5]{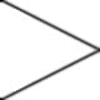}} & \multicolumn{1}{c|}{\includegraphics[scale=0.5]{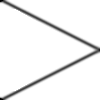}} & \multicolumn{1}{c|}{\includegraphics[scale=0.5]{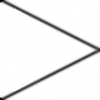}} & \multicolumn{1}{c|}{\includegraphics[scale=0.5]{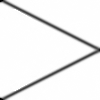}} & \multicolumn{1}{c|}{\includegraphics[scale=0.5]{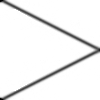}} & \multicolumn{1}{c|}{\includegraphics[scale=0.5]{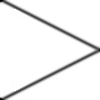}} & \\
			\cline{3-8}
			& & \multicolumn{6}{c|}{$\{V_\text{err}$ , $Obj_\text{err}\}$ (\%)} & \\
			\cline{3-8}
			& & $\{7.4, 3.8263\}$ & $\{12.4, 0.1351\}$ & $\{2.2, 10.1873\}$ & $\{6, 2.67\}$ & $\{3.4, 2.9814\}$ & $\{0.6, 2.8736\}$ & \\
			\hline
			\multirow{6}{*}{0.15} & \multirow{6}{*}{\includegraphics[scale=0.075]{vf_15_in.png}} & \multicolumn{1}{c|}{\multirow{4}{*}{\includegraphics[scale=0.5]{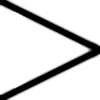}}} & \multicolumn{1}{c|}{\multirow{4}{*}{\includegraphics[scale=0.5]{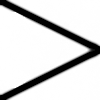}}} & \multicolumn{1}{c|}{\multirow{4}{*}{\includegraphics[scale=0.5]{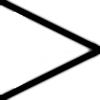}}} & \multicolumn{1}{c|}{\multirow{4}{*}{\includegraphics[scale=0.5]{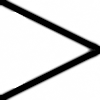}}} & \multicolumn{1}{c|}{\multirow{4}{*}{\includegraphics[scale=0.5]{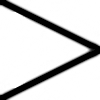}}} & \multirow{4}{*}{\includegraphics[scale=0.5]{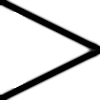}} & \multirow{6}{*}{\includegraphics[scale=0.5]{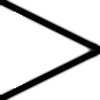}} \\
			& & \multicolumn{1}{c|}{} & \multicolumn{1}{c|}{} & \multicolumn{1}{c|}{} & \multicolumn{1}{c|}{} & \multicolumn{1}{c|}{} & \multicolumn{1}{c|}{} & \\
			& & \multicolumn{1}{c|}{} & \multicolumn{1}{c|}{} & \multicolumn{1}{c|}{} & \multicolumn{1}{c|}{} & \multicolumn{1}{c|}{} & \multicolumn{1}{c|}{} & \\
			& & \multicolumn{1}{c|}{} & \multicolumn{1}{c|}{} & \multicolumn{1}{c|}{} & \multicolumn{1}{c|}{} & \multicolumn{1}{c|}{} & \multicolumn{1}{c|}{} & \\
			\cline{3-8}
			& & \multicolumn{6}{c|}{$\{V_\text{err}$ , $Obj_\text{err}\}$ (\%)} & \\
			\cline{3-8}
			& & $\{1.67, 0.0356\}$ & $\{0.867, 0.1048\}$ & $\{0, 0.7071\}$ & $\{1.73, 0.7150\}$ & $\{0.067, 0.1337\}$ & $\{0.53, 0.4479\}$ & \\
			\hline
			\multirow{6}{*}{0.25} & \multirow{6}{*}{\includegraphics[scale=0.075]{vf_25_in_final.png}} & \multicolumn{1}{c|}{\multirow{4}{*}{\includegraphics[scale=0.5]{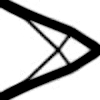}}} & \multicolumn{1}{c|}{\multirow{4}{*}{\includegraphics[scale=0.5]{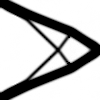}}} & \multicolumn{1}{c|}{\multirow{4}{*}{\includegraphics[scale=0.5]{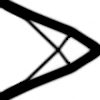}}} & \multicolumn{1}{c|}{\multirow{4}{*}{\includegraphics[scale=0.5]{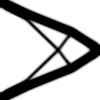}}} & \multicolumn{1}{c|}{\multirow{4}{*}{\includegraphics[scale=0.5]{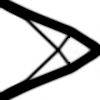}}} & \multirow{4}{*}{\includegraphics[scale=0.5]{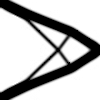}} & \multirow{6}{*}{\includegraphics[scale=0.5]{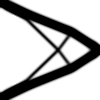}} \\
			& & \multicolumn{1}{c|}{} & \multicolumn{1}{c|}{} & \multicolumn{1}{c|}{} & \multicolumn{1}{c|}{} & \multicolumn{1}{c|}{} & \multicolumn{1}{c|}{} & \\
			& & \multicolumn{1}{c|}{} & \multicolumn{1}{c|}{} & \multicolumn{1}{c|}{} & \multicolumn{1}{c|}{} & \multicolumn{1}{c|}{} & \multicolumn{1}{c|}{} & \\
			& & \multicolumn{1}{c|}{} & \multicolumn{1}{c|}{} & \multicolumn{1}{c|}{} & \multicolumn{1}{c|}{} & \multicolumn{1}{c|}{} & \multicolumn{1}{c|}{} & \\
			\cline{3-8}
			& & \multicolumn{6}{c|}{$\{V_\text{err}$ , $Obj_\text{err}\}$ (\%)} & \\
			\cline{3-8}
			& & $\{0.88, 0.2312\}$ & $\{0.4, 0.2615\}$ & $\{0.16, 0.9801\}$ & $\{0.36, 0.6686\}$ & $\{0.44, 0.4134\}$ & $\{0.24, 0.3365\}$ & \\
			\hline
			\multirow{6}{*}{0.40} & \multirow{6}{*}{\includegraphics[scale=0.075]{vf_40_in.png}} & \multicolumn{1}{c|}{\multirow{4}{*}{\includegraphics[scale=0.5]{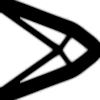}}} & \multicolumn{1}{c|}{\multirow{4}{*}{\includegraphics[scale=0.5]{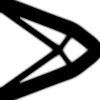}}} & \multicolumn{1}{c|}{\multirow{4}{*}{\includegraphics[scale=0.5]{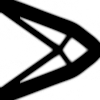}}} & \multicolumn{1}{c|}{\multirow{4}{*}{\includegraphics[scale=0.5]{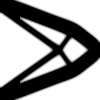}}} & \multicolumn{1}{c|}{\multirow{4}{*}{\includegraphics[scale=0.5]{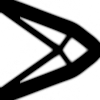}}} & \multirow{4}{*}{\includegraphics[scale=0.5]{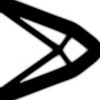}} & \multirow{6}{*}{\includegraphics[scale=0.5]{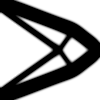}} \\
			& & \multicolumn{1}{c|}{} & \multicolumn{1}{c|}{} & \multicolumn{1}{c|}{} & \multicolumn{1}{c|}{} & \multicolumn{1}{c|}{} & \multicolumn{1}{c|}{} & \\
			& & \multicolumn{1}{c|}{} & \multicolumn{1}{c|}{} & \multicolumn{1}{c|}{} & \multicolumn{1}{c|}{} & \multicolumn{1}{c|}{} & \multicolumn{1}{c|}{} & \\
			& & \multicolumn{1}{c|}{} & \multicolumn{1}{c|}{} & \multicolumn{1}{c|}{} & \multicolumn{1}{c|}{} & \multicolumn{1}{c|}{} & \multicolumn{1}{c|}{} & \\
			\cline{3-8}
			& & \multicolumn{6}{c|}{$\{V_\text{err}$ , $Obj_\text{err}\}$ (\%)} & \\
			\cline{3-8}
			& & $\{0.75, 0.0852\}$ & $\{0.275, 0.0654\}$ & $\{0.2, 0.1345\}$ & $\{0.4, 0.2571\}$ & $\{0.175, 0.0648\}$ & $\{0.125, 0.2143\}$ & \\
			\hline
			\multirow{6}{*}{0.60} & \multirow{6}{*}{\includegraphics[scale=0.075]{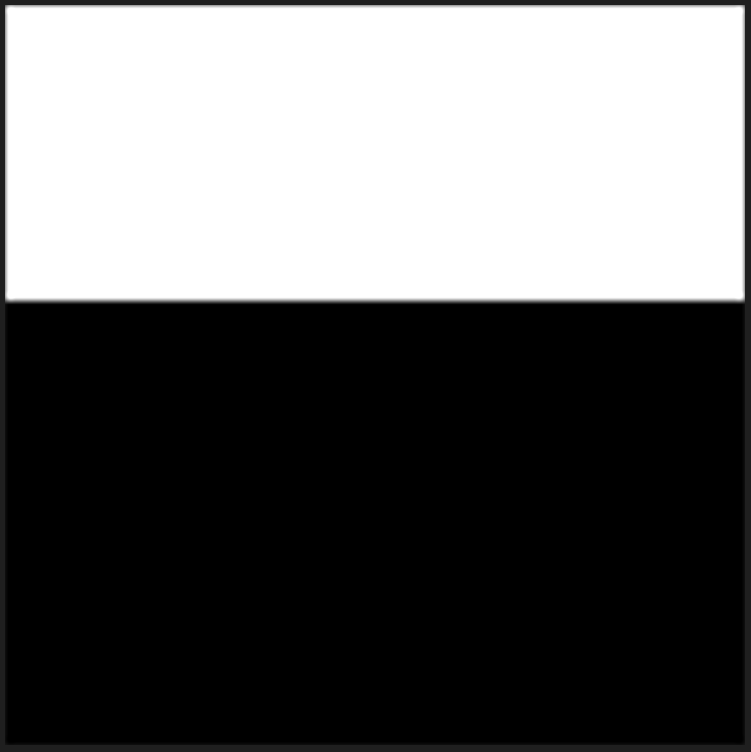}} & \multicolumn{1}{c|}{\multirow{4}{*}{\includegraphics[scale=0.5]{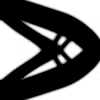}}} & \multicolumn{1}{c|}{\multirow{4}{*}{\includegraphics[scale=0.5]{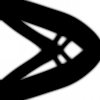}}} & \multicolumn{1}{c|}{\multirow{4}{*}{\includegraphics[scale=0.5]{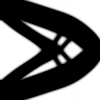}}} & \multicolumn{1}{c|}{\multirow{4}{*}{\includegraphics[scale=0.5]{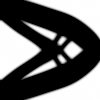}}} & \multicolumn{1}{c|}{\multirow{4}{*}{\includegraphics[scale=0.5]{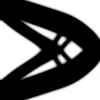}}} & \multirow{4}{*}{\includegraphics[scale=0.5]{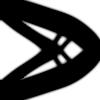}} & \multirow{6}{*}{\includegraphics[scale=0.5]{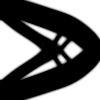}} \\
			& & \multicolumn{1}{c|}{} & \multicolumn{1}{c|}{} & \multicolumn{1}{c|}{} & \multicolumn{1}{c|}{} & \multicolumn{1}{c|}{} & \multicolumn{1}{c|}{} & \\
			& & \multicolumn{1}{c|}{} & \multicolumn{1}{c|}{} & \multicolumn{1}{c|}{} & \multicolumn{1}{c|}{} & \multicolumn{1}{c|}{} & \multicolumn{1}{c|}{} & \\
			& & \multicolumn{1}{c|}{} & \multicolumn{1}{c|}{} & \multicolumn{1}{c|}{} & \multicolumn{1}{c|}{} & \multicolumn{1}{c|}{} & \multicolumn{1}{c|}{} & \\
			\cline{3-8}
			& & \multicolumn{6}{c|}{$\{V_\text{err}$ , $Obj_\text{err}\}$ (\%)} & \\
			\cline{3-8}
			& & $\{0.25, 0.0252\}$ & $\{0.183, 0.0092\}$ & $\{0.033, 0.013\}$ & $\{0.183, 0.0788\}$ & $\{0.067, 0\}$ & $\{0.017, 0.025\}$ & \\
			\hline
			\multirow{6}{*}{0.65} & \multirow{6}{*}{\includegraphics[scale=0.075]{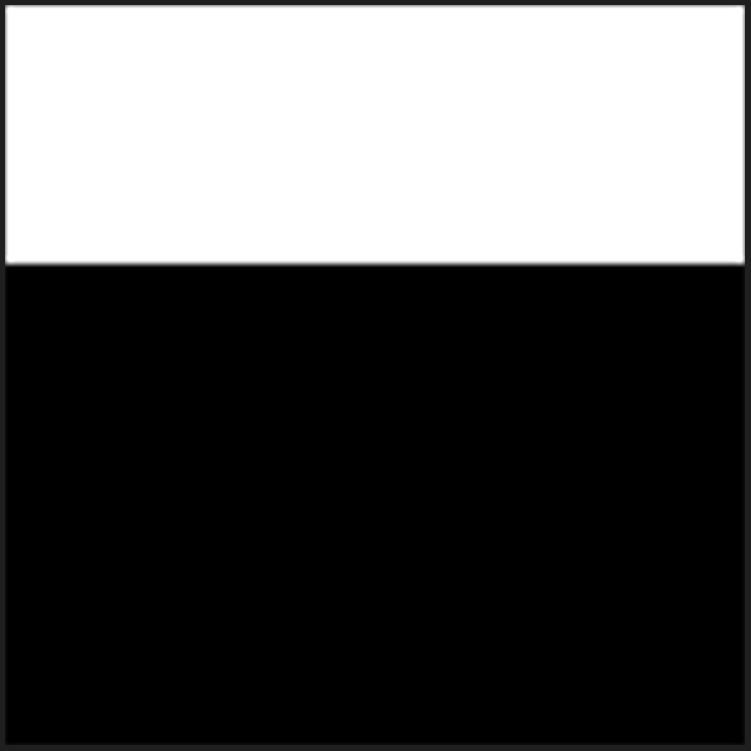}} & \multicolumn{1}{c|}{\multirow{4}{*}{\includegraphics[scale=0.5]{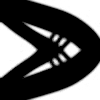}}} & \multicolumn{1}{c|}{\multirow{4}{*}{\includegraphics[scale=0.5]{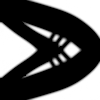}}} & \multicolumn{1}{c|}{\multirow{4}{*}{\includegraphics[scale=0.5]{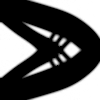}}} & \multicolumn{1}{c|}{\multirow{4}{*}{\includegraphics[scale=0.5]{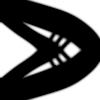}}} & \multicolumn{1}{c|}{\multirow{4}{*}{\includegraphics[scale=0.5]{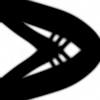}}} & \multirow{4}{*}{\includegraphics[scale=0.5]{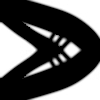}} & \multirow{6}{*}{\includegraphics[scale=0.5]{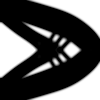}} \\
			& & \multicolumn{1}{c|}{} & \multicolumn{1}{c|}{} & \multicolumn{1}{c|}{} & \multicolumn{1}{c|}{} & \multicolumn{1}{c|}{} & \multicolumn{1}{c|}{} & \\
			& & \multicolumn{1}{c|}{} & \multicolumn{1}{c|}{} & \multicolumn{1}{c|}{} & \multicolumn{1}{c|}{} & \multicolumn{1}{c|}{} & \multicolumn{1}{c|}{} & \\
			& & \multicolumn{1}{c|}{} & \multicolumn{1}{c|}{} & \multicolumn{1}{c|}{} & \multicolumn{1}{c|}{} & \multicolumn{1}{c|}{} & \multicolumn{1}{c|}{} & \\
			\cline{3-8}
			& & \multicolumn{6}{c|}{$\{V_\text{err}$ , $Obj_\text{err}\}$ (\%)} & \\
			\cline{3-8}
			& & $\{0.23, 0.007\}$ & $\{0.1076, 0.067\}$ & $\{0.77, 0.094\}$ & $\{0.139, 0.105\}$ & $\{0.1692, 0.056\}$ & $\{0.077, 0.044\}$ & \\
			\hline
			\multirow{6}{*}{0.75} & \multirow{6}{*}{\includegraphics[scale=0.075]{vf_75_in_final.png}} & \multicolumn{1}{c|}{\multirow{4}{*}{\includegraphics[scale=0.5]{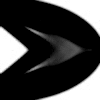}}} & \multicolumn{1}{c|}{\multirow{4}{*}{\includegraphics[scale=0.5]{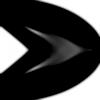}}} & \multicolumn{1}{c|}{\multirow{4}{*}{\includegraphics[scale=0.5]{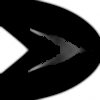}}} & \multicolumn{1}{c|}{\multirow{4}{*}{\includegraphics[scale=0.5]{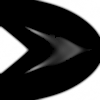}}} & \multicolumn{1}{c|}{\multirow{4}{*}{\includegraphics[scale=0.5]{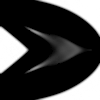}}} & \multirow{4}{*}{\includegraphics[scale=0.5]{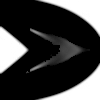}} & \multirow{6}{*}{\includegraphics[scale=0.5]{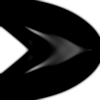}} \\
			& & \multicolumn{1}{c|}{} & \multicolumn{1}{c|}{} & \multicolumn{1}{c|}{} & \multicolumn{1}{c|}{} & \multicolumn{1}{c|}{} & \multicolumn{1}{c|}{} & \\
			& & \multicolumn{1}{c|}{} & \multicolumn{1}{c|}{} & \multicolumn{1}{c|}{} & \multicolumn{1}{c|}{} & \multicolumn{1}{c|}{} & \multicolumn{1}{c|}{} & \\
			& & \multicolumn{1}{c|}{} & \multicolumn{1}{c|}{} & \multicolumn{1}{c|}{} & \multicolumn{1}{c|}{} & \multicolumn{1}{c|}{} & \multicolumn{1}{c|}{} & \\
			\cline{3-8}
			& & \multicolumn{6}{c|}{$\{V_\text{err}$ , $Obj_\text{err}\}$ (\%)} & \\
			\cline{3-8}
			& & $\{0.027, 0.112\}$ & $\{0.013, 0.022\}$ & $\{0.093, 0.112\}$ & $\{0.13, 0.097\}$ & $\{0, 0.011\}$ & $\{0.347, 0.343\}$ & \\
			\hline
	\end{tabular}}
\end{table*}
\begin{figure*}
	\centering
	\begin{subfigure}{0.30\textwidth}
		\centering
		\includegraphics[scale = 0.6]{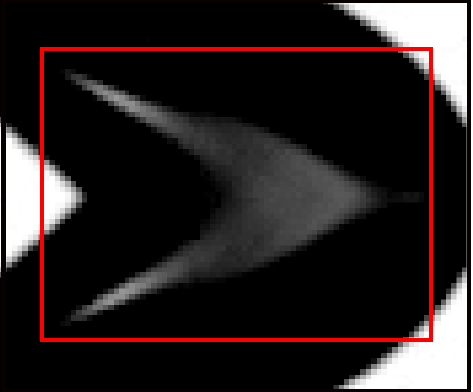}
		\caption{$n = 0$}
		\label{fig:mid_pred_0_vf_75_error}
	\end{subfigure}
	\quad
	\centering
	\begin{subfigure}{0.30\textwidth}
		\centering
		\includegraphics[scale = 0.6]{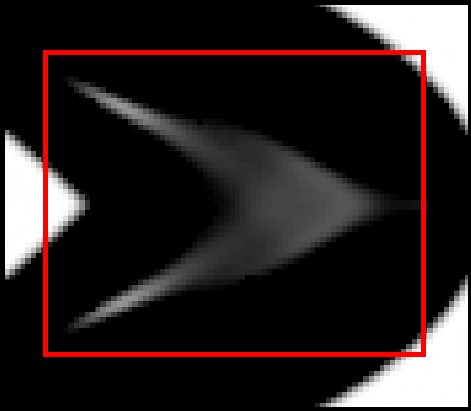}
		\caption{$n = 8000$}
		\label{fig:mid_pred_8000_vf_75_error}
	\end{subfigure}
	\quad
	\begin{subfigure}{0.30\textwidth}
		\centering
		\includegraphics[scale = 0.6]{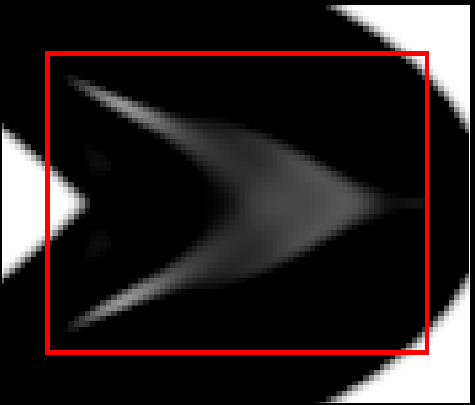}
		\caption{Target design}
		\label{fig:mid_targ_75}
	\end{subfigure}
	\caption{Optimized design comparison}
	\label{fig:comparison_mid_adap}
\end{figure*}
\subsubsection{Cantilever beam with end load}
We next solve the cantilever beam with an end load problem (Fig.~\ref{fig:can_1end}). The design domain, boundary conditions, and applied load are shown in Fig.~\ref{fig:can_1end}.  
Like the mid-load problem, we generated a total of 95 training examples using \ttt{top88} MATLAB code (Sec.~\ref{Sec:GentrainData}) and trained the CNN architecture for 2000 epochs. 

Different volume fractions are used to produce the CNN results, as shown in Table~\ref{Table:CNNresults2}. Once the model is trained, the results are obtained in a fraction of a second. One notes that the optimized designs obtained by the proposed CNN model closely topologically resemble those obtained using the MATLAB code. The model generated optimized designs having low volume fraction and objective function value errors (Table~\ref{Table:CNNresults2}). As noted in the mid-load problem, negligible deviations in the feature shape and size with insignificant effects on the performance can be noted. For example, the result obtained with $V_f = 0.70$,  is very close to the target with little deviation marked in red rectangular, see Fig.~\ref{fig:comparison_bot}. Corresponding $V_\text{err}$ and $Obj_\text{err}$ are 0.19\% and 0.063\%, which are negligible.  

\PKnew{The used volume fractions in Table~\ref{Table:CNNresults1} and Table~\ref{Table:CNNresults2} are chosen arbitrary for generating the CNN results, though they are available in the training dataset. Note that the trained neural network is not just regurgitating designs from the dataset and is producing these outputs by learning the relationships within the provided training data; that is why $V_\text{err}$ and $obj_\text{err}$ are noted. We provide results with a set of different volume fractions not present in the training dataset. This is to emphasize that once the proposed CNN model is trained, it produces results for any volume fraction in a fraction of a second.}

\PKnew{Table~\ref{Table:CNNresults3c} shows results for volume fractions (not present in the training dataset) 17.5\%, 36.6\%, 44.7\%, 55.8\%, and 72.1\%. The associated $V_\text{err}$ and $obj_\text{err}$ are also provided. The topologies of the CNN results exactly resemble those of their target designs. The maximum $V_\text{err} = 2.17\%$ is noted for $V_f = 17.5\%$; however, negligible $obj_\text{err} = 0.49\%$ is noted, as topologies obtained via CNN model and target resemble closely. For this case, too, $V_\text{err}$ and $obj_\text{err}$ are relatively less, indicating the robustness of the proposed CNN model.}

\subsubsection*{\PKnew{Influence of number of the training set}}

\PKnew{Next, we demonstrate the capability of the proposed CNN model to produce decent results even with a few training datasets. We train the proposed CNN architecture with 32 training samples (volume fractions ranging from 0.02:0.03:0.95). Table~\ref{Table:CNNresults3d} displays the results for volume fractions 15\%, 25\%, 40\%, 60\%, 70\%, and 85\%. These volume fractions are not a part of the provided training dataset. Results in Table~\ref{Table:CNNresults3d} have relatively higher $V_{err}$  and $Obj_{err}$ compared with Table~\ref{Table:CNNresults2}, which is obvious since the former is trained with 32 training datasets, whereas the latter is trained with 95 training datasets. However, the topologies of the CNN results closely resemble those with target designs.}

\subsubsection{Design-dependent pressure loadbearing arch structure}
To demonstrate the versatility of the proposed model, a design-dependent pressure loadbearing arch structure is optimized here. The design domain, boundary conditions, and applied pressure loads are shown in Fig.~\ref{fig:archDD}. 
95 training examples are generated using \ttt{TOPress} MATLAB code~\citep{kumar2023TOPress}. The CNN architecture is trained using the generated data for 2000 epochs. 

Results obtained using the base CNN model are depicted in Table.~\ref{Table:CNNresults_topress} for different volume fractions. The trained model takes a fraction of a second to produce the optimized results. Dealing with design-dependent pressure loads in TO has several unique challenges, as the direction, magnitude, and/or location get updated with TO evolution~\citep{kumar2020topology,kumar2023TOPress}. However, once the neural network model is trained for the specific problem, the user naturally gets rid of such challenges and generates the optimized designs in a fraction of a second. 

The model generates correct optimized designs as $V_\text{err}$ and $Obj_\text{err}$ are within the acceptable range, and they topologically resemble those obtained using \ttt{TOPress}~\citep{kumar2023TOPress}. Though the problem is challenging in nature~\citep{kumar2020topology,kumar2023TOPress}, its optimized design lacks complex patterns. For such problems with simple optimized designs, the proposed base CNN model is sufficient to capture the topology accurately.

\subsubsection{Bulk Modulus optimization problem}
Having demonstrated the compliance minimization problem with and without design-dependent loads using the proposed CNN model, this section provides a material bulk modulus optimization problem (Fig.~\ref{fig:toxDD}). We generate 68 training examples by varying the volume fraction from 0.03 to 0.70 with an increment of 0.01 using \ttt{topX} MATLAB code~\citep{xia2015design}. The training data used for this case is relatively low due to limitations of  \ttt{topX} MATLAB code. Using the data, the model is trained with 2000 epochs.

The results obtained using the CNN model and the MATLAB code are depicted in Table~\ref{Table:CNNresults_topX}. The results obtained using the former topologically resemble those obtained from the latter. However, one notices relatively high $V_\text{err}$ compare  to previous instances (Table ~\ref{Table:CNNresults_topX}).

\subsection{Results with the adaptive layer}\label{Sec:ResWAdpLay}
This section presents solutions to a cantilever beam problem with center load (Fig.~\ref{fig:can_1cent}) using the model with an adaptive layer in the dense layer~(Fig.~\ref{fig:CNNarchitech}). The layer is added by uncommenting line~38. Using this example, we go ahead and explain how the optimized designs generated by the model can be improved by the addition of the adaptive layer.
An additional adaptive layer with $n$ neurons can help capture intricate image patterns as the number of learning parameters increases in the CNN model. We generate 95 training examples using the same procedure employed previously. In addition to training the \textit{base} architecture,  we train different networks with different $n$ for the adaptive layer with 2000 epochs.

Table~\ref{Table:CNNresults_mid_adap} displays the results obtained for different volume fractions with various $n$ in the adaptive layer. We also include $\{V_\text{err},\,Obj_\text{err}\}$ corresponding to each solution. One notes that given the measure $\{V_\text{err},\,Obj_\text{err}\}$ and a visual inspection, the CNN provides the highest quality results with $n=8000$. A close look for results with a volume fraction of 0.75 is displayed in Fig.~\ref{fig:comparison_mid_adap}.

\section{Concluding remarks}\label{Sec:conc}
This paper proposes an adaptive deep Convolutional Neural Network architecture to tackle the multidisciplinary TO problems. The model employs an encoder-decoder type architecture with dense layers introduced to help capture the complex geometrical features for the optimized designs. The incorporation of the adaptive layer introduces flexibility to the initially rigid architecture, granting users a degree of control over the outputs generated by the model. Publicly available MATLAB codes are used to generate the data for training purposes. The efficacy and success of the developed CNN architecture are tested by generating optimized designs for compliance minimization problems with constant and design-dependent loads, and also on a material bulk modulus maximization problem. Once the network is trained with a certain number of epochs, it gives the sought-after optimized designs instantaneously.

Despite the slight inaccuracies observed in the outputs generated by the CNN when intricate patterns are present in optimized designs, the introduced deep learning architecture represents a valuable endeavor to leverage machine learning for expediting the TO process. In contrast to prior automation efforts of TO problems using convolutional neural networks, our approach yields remarkably precise outcomes while requiring significantly less training data.

The proposed model has a fixed domain of $100\times100$, and output (optimized designs) can be generated explicitly for the boundary and force conditions for which the training data has been provided. Tapping the power of the deep neural network, generalizing the proposed network for design domain size, and obtaining the output results for the boundary and force conditions that are not used for the training data open up exciting avenues for further research.

The paper also provides the Python code, called \ttt{PyTOCNN}, for the proposed CNN model in Appendix~\ref{Sec:PyTOaCNN}. Each part of the code is explained in detail. \ttt{PyTOCNN} is included to facilitate the reproducibility of the presented results. Additionally, we believe that \ttt{PyTOCNN} can provide a potential gateway and tool for students and newcomers in the field of TO with machine learning.

\begin{appendices}
	\onecolumn
	\section{\ttt{PyTOaCNN}: CNN topology optimization Python code}\label{Sec:PyTOaCNN}
\begin{lstlisting}[language=Python, basicstyle=\footnotesize\ttfamily,breaklines=true]
#=============== PyTOaCNN: A CNN topology optimization Python code ========
#================== Installing Tensorflow and importing libraries =========
pip install tensorflow
import os
import numpy as np
import matplotlib.pyplot as plt
import tensorflow as tf 
from matplotlib.image import imread
from tensorflow.keras.models import Sequential
from tensorflow.keras.layers import Conv2D, Conv2DTranspose, MaxPool2D
#========================= Providing training data file path ===============
Input_train_folder_path ='main\input_data'
Output_train_folder_path ='main\output_data'
Input_train_elements = os.listdir(Input_train_folder_path) 
Output_train_elements = os.listdir(Output_train_folder_path)
#======================= Developing the input and output training tensors ==
Input_train = np.zeros((95,100,100,1))
Output_train= np.zeros((95,100,100,1))
for index, Input_train_element in enumerate(Input_train_elements):
element_path = os.path.join(Input_train_folder_path, Input_train_element)
img = imread(element_path)
img = img.reshape((100, 100, 1))
Input_train[index] = img   
for index, Output_train_element in enumerate(Output_train_elements):
element_path = os.path.join(Output_train_folder_path, Output_train_element)
img = imread(element_path)
img = img.reshape((100, 100, 1))
Output_train[index] = img
#========================== Developing the CNN Model ========================
model = Sequential()
model.add(Conv2D(filters=128,kernel_size=(2,2),strides=(1,1),padding='same',input_shape=(100,100,1),activation='relu'))
model.add(MaxPool2D(pool_size=(2,2),strides=(2,2)))
model.add(Conv2D(filters= 256,kernel_size=(2,2),strides=(1,1),padding='same',activation='relu'))
model.add(MaxPool2D(pool_size=(2,2),strides=(2,2)))
model.add(Conv2D(filters= 512,kernel_size=(5,5),strides=(1,1),padding='same',activation='relu'))
model.add(MaxPool2D(pool_size=(5,5),strides=(5,5)))
model.add(tf.keras.layers.Flatten())
model.add(tf.keras.layers.Dense(units= n , activation='relu')) # Provides values for n
model.add(tf.keras.layers.Dense(units=12800, activation='relu'))          
model.add(tf.keras.layers.Reshape(target_shape=(5,5,512)))         
model.add(Conv2DTranspose(filters=256,kernel_size=(2,2),strides=(2,2),activation='relu'))
model.add(Conv2DTranspose(filters=128,kernel_size=(5,5),strides=(5,5),activation='relu'))
model.add(Conv2DTranspose(filters=1,kernel_size=(2,2),strides=(2,2),activation='relu'))
model.compile(optimizer='adam', loss='mean_squared_error')
#===========================Training command===================================
model.summary()
model.fit(Input_train,Output_train,epochs = 2000)
\end{lstlisting}

\end{appendices}


\end{document}